\documentclass[11pt]{article}
\usepackage[affil-it]{authblk}

\usepackage{color}
\usepackage{amsmath,amsfonts,amssymb,mathrsfs,amsthm}
\usepackage{graphicx}
\usepackage{listings}
\usepackage{enumerate,enumitem}
\usepackage{mathtools}
\usepackage[a4paper, total={6in, 8in}]{geometry}
\usepackage{cite}
\usepackage[colorlinks,linkcolor=blue,citecolor=blue]{hyperref}
\usepackage{empheq}
\usepackage{pdfpages}
\usepackage[titletoc]{appendix}
\usepackage[utf8]{inputenc}
\usepackage{xspace}
\numberwithin{equation}{section}

\newcommand{\Eqref}[1]{Eq.~\eqref{#1}}
\newcommand{\Eqsref}[1]{Eqs.~\eqref{#1}}
\newcommand{\e}[1]{\text{e}^{#1}}

\newcommand{\Figref}[1]{Fig.~\ref{#1}}
\newcommand{\Sectionref}[1]{Section~\ref{#1}}
\newcommand{\subfig}[2]{ Fig.~\hyperref[#1]{\ref{#1}#2}}

\newcommand{\T}{\mathcal{T}}

\newtheorem{Result}{Result}

\newcounter{mnotecount}[section]
\let\oldmarginpar\marginpar
\renewcommand\marginpar[1]{\-\oldmarginpar[\raggedleft\footnotesize #1]%
	{\raggedright\footnotesize #1}}

\title{Non-variational scalar field cosmology: Exact Bianchi I solutions for near-minimal scalar fields}
\author{J. Ritchie\footnote{Email: josh.ritchie@otago.ac.nz}}
\affil{Department of Mathematics and Statistics, University of Otago, New Zealand.}

\begin{document}
	\maketitle
	\begin{abstract}
		The purpose of this work is to investigate spatially homogeneous and flat cosmological solutions of the Einstein equations coupled to a non-variational ``near-minimal'' scalar field. This coupling model represents a minimal departure from standard theory by decoupling the scalar field's self-interaction term from the derivative of its potential. By assuming a quadratic potential and a self-interaction term that is proportional to the potential, we derive four new exact Bianchi I solutions. We demonstrate that these solutions produce a diverse range of cosmological phenomena, including Big Bang, Big Crunch, and Big Rip singularities, as well as oscillatory (``cyclic'') behaviour. For our exact solutions, these singularities occur in infinite proper time and hence are never truly reachable by an observer. To assess the stability of these cosmologies, we perform a numerical stability analysis against spatially inhomogeneous perturbations of the mean curvature. We find that the oscillatory solution is unstable to perturbations of this type, as are solutions in possession of a crushing singularity. Conversely, solutions with a Big Rip singularity (at infinity) are stable to spatially inhomogeneous perturbations of the mean curvature. 
	\end{abstract}
	\section{Introduction}
	\paragraph{Motivation.}	Exact solutions --of the Einstein equations-- play a key role in General Relativity (GR) and cosmology \cite{ExactSolutions:Stephani}. To date, exact solutions have been exclusively derived under the assumptions of high degrees of symmetry \cite{ExactSolutions:Stephani,EmptySpacetimes:Taub,BeyerHennig:2012,BeyerHennig:2014}. For example, the Schwarzschild spacetime --which provides the simplest known description of a black hole-- is derived under the assumption of spherical symmetry \cite{Schwarzschild:1916,Gravitation}. Despite this, exact solutions nevertheless provide researchers with models that can be used to help explain the dynamics of more general classes of solutions \cite{GriffithsPodolsky:2009}.  
	
	Of particular interest for the present work is the \emph{cosmological} setting. Here, the prototypical solutions are the Friedman-Lema\^{i}tre-Robertson-Walker (FLRW) spacetimes, which are a class of spatially homogeneous and isotropic cosmologies that are generally used as models of our present universe \cite{Friedmann:1922,Lemaitre:1927,Robertson:1935,Walker:1937}.  
	
	However, observations suggest that our universe is not perfectly isotropic and as such it is also of interest to study classes of spacetimes that are still spatially homogeneous but are \emph{not} isotropic (i.e., anisotropic) \cite{PlanckIsotropy:2020,CosmologicalPrincipleReview:Aluri}. In this setting, spatially homogeneous solutions of the Einstein+matter equations are commonly categorised using the \emph{Bianchi classification} \cite{Bianchi:1898,WainwrightEllis1997}. Formally, these spacetimes are defined by the presence of three (spatial) Killing vectors (which describe the symmetry properties of the spacetime). The various types of Bianchi spacetimes are defined based on the structure of the commutator of said Killing vectors \cite{WainwrightEllis1997}. The simplest type of spatially homogeneous solution is the class of Bianchi I cosmologies, for which the pairwise commutator of the Killing vectors is identically zero. Geometrically, this ends up implying that Bianchi I solutions are spatially flat \cite{BianchiClass:EllisMacCallum}. The prototypical Bianchi I solutions (of the Einstein vacuum equations) are the \emph{Kasner} spacetimes, which are a one-parameter family of anisotropic cosmologies \cite{KasnerMetric}. These solutions have found significant application in mathematical cosmology, largely due to their role in questions related to stable Big Bang formation \cite{RodnianskiSpeck:Linear,RodnianskiSpeck:NonLinear,BeyerOliynyk:2023_1,BeyerOliynyk:2023_2}. 
	
	\paragraph{Scalar field models.} While the vacuum Bianchi I solutions (of the Einstein equations) are undeniably important, they lack the energy density required to drive the expansion observed in our universe. In particular, they are unable to account for the matter- or radiation-dominated phases of cosmic history \cite{WainwrightEllis1997}. To account for these effects it is common to consider the addition of a \emph{scalar field}, which provides a dynamical mechanism responsible for the observed acceleration of our universe \cite{Guth:1981}. 
	
	There are several different types of scalar field appearing in the literature, each of which has been introduced to solve different problems. For example, a minimally coupled scalar field --which is commonly referred to as the ``inflaton''-- has been introduced to provide a dynamical mechanism responsible for generating a period of accelerated expansion in the early universe commonly referred to as ``inflation'' \cite{Guth:1981}. Inflation itself was introduced as a solution to the so-called \emph{horizon problem} (see, for example, \cite{Guth:1981,Liddle:1999,Sloan:2007}). Similarly, non-canonical $k$-essence scalar field models can be used to dynamically ``track'' the matter content of the universe and have been introduced as a possible solution to the \emph{cosmic coincidence problem} \cite{Garriga:1999,Vikman:2007,Armendariz-Picon:2001,Babichev:2008}.
	
	The vast majority of scalar field models appearing in the literature are derived from a variational principle (e.g., the minimally coupled scalar field). In this setting one starts with a Lagrangian from which both the scalar field's equations of motion and the corresponding energy-momentum tensor are derived. In the case of a minimally coupled scalar field, this leads to exactly one degree of freedom appearing in the resulting equations --namely, the potential. It is unclear exactly what form the potential ``should'' take, although there are several works exploring the various sorts of conditions one might expect it to satisfy. For example, in the future time direction, the so-called slow-roll conditions are necessary for a smooth exit from inflation \cite{SlowRoll:Linde,Beyer:2013}. Conversely, in the past time direction (towards the initial singularity) works such as \cite{KasnerSolutions,Ritchie2022} establish necessary conditions on the potential that ensure the corresponding scalar field is \emph{Kasner-like}. 
	
	While these sorts of conditions constrain the possible choices of scalar field potential, the exact functional form of the potential remains unclear, and different choices can lead to very different qualitative behaviours. On the one hand, works such as \cite{Koutvitsky:2005,Koutvitsky:2006,Drees:2025} considered potentials that lead to \emph{oscillatory} universes. Here, the volume of the universe grows until it reaches a maximal value and then contracts back to its initial size. On the other hand, works such as \cite{kujat2006phantom,chimento2004big,Briscese:2007} have found universes that possess a \emph{Big Rip}. This is a type of singularity in which the volume of the universe becomes infinite near the singular point. This contrasts typical \emph{crushing} type singularities (such as the Big Bang) for which the volume of the universe is infinitely small near the singular point. All of these results have been established for variational type scalar field couplings. Although less common, non-variational models have also appeared in the literature \cite{Gao:2010,Ritchie:2026}. Such non-variational models may provide a useful framework for describing non-conservative dynamics, such as those appearing in modified gravity theories (for a review on modified gravity we refer the interested reader to \cite{Clifton:2012}) or higher-dimensional brane-world scenarios (see, for example, \cite{Maartens:2010} and the references therein). Establishing a precise connection between these theories and the specific non-variational framework of \cite{Ritchie:2026} remains a subject for future investigation. In addition to non-variational scalar field models (which are the focus of this work here) it is worth briefly noting that there are other types of non-standard matter (e.g., thermodynamic fluids \cite{Israel:1979,Maartens:1995,Maartens:1996}, and, interacting dark-energy-dark-matter models \cite{Honorez:2010,VanDerWesthuizen:2024}), which share with the present work the property of not being straightforwardly derivable from a single Lagrangian.
	
	\paragraph{Goals of the current work.} Of particular interest here is the recent work, by us, in \cite{Ritchie:2026}, where we introduced a generic framework for coupling scalar fields to gravity without the use of a variational principle. Crucially, standard scalar field couplings (such as the minimal and $k$-essence scalar fields) could be recovered within our framework and hence correspond to specific models within a more general class of permissible scalar field couplings.	In addition to presenting our framework, we also investigated a type of scalar field coupling that we referred to as ``near-minimal'': Here, all choices (available in our framework) were chosen to match the standard minimal scalar field with the exception of the self-interaction term in the scalar field's equation of motion --which was allowed to differ from the standard case where it is equal to the derivative of the potential. This ``near-minimal'' coupling represents the smallest possible departure from the standard minimally coupled scalar field, providing a clear baseline for how non-variational terms modify cosmological evolution. In this setting, we established  conditions under which these near-minimal models could be asymptotically matched to a Kasner solution, near the initial singularity.
	
	Given these results, it is now of interest to investigate the types of solutions one might expect near-minimal scalar fields to produce. This is exactly the goal of the present work. In particular, we focus on constructing exact Bianchi I solutions of the near-minimal system. The results presented here should be understood as exploratory in nature. Indeed, several of the choices made throughout this work have been made for mathematical convince only. Although the physical consequences of these choices is of course interesting, it is not our focus here and shall not be discussed. The specific aim of this work is to investigate the range of solutions possible within this near-minimal coupling and to investigate their stability properties.  To that end, we construct these exact solutions under the restriction that the scalar field's self-interaction term is proportional to the potential function (and, in particular, is \emph{not} equal to the derivative of the potential). The potential itself is chosen to be a quadratic function. Our first main finding is summarized as follows.
	\begin{Result}
		We construct four exact Bianchi I solutions that exhibit known cosmological phenomena, such as Big Rip, Big Bang, and Big Crunch singularities as well as oscillatory behaviour and an asymptotically static universe.
	\end{Result}
	As stated above, we do not expect these solutions to be physically relevant models of our universe. Indeed, for many of these solutions the weak null energy condition is not satisfied. For this reason we do not provide a detailed investigation of their physics. We do, however, discuss the long-term dynamics of these solutions. Note here that, for each solution in possession of a singularity, the singularity itself occurs in infinite proper time and as such is never truly reachable by an observer. We then assess the stability properties of these solutions under mean curvature perturbations. 
	
	For spatially homogenous mean curvature perturbations we use a mixture of analytical and numerical techniques to provide a detailed description the solutions behaviour under perturbation. Our findings can be summarised as follows:
	\begin{Result}
		For spatially homogenous mean curvature perturbations, the oscillatory solution is stable to sufficiently small perturbations of the mean curvature. Moreover, all solutions in possession of a crushing singularity are unstable to perturbation, with the singularity occurring in finite proper time. Conversely, all solutions in possession of a Big Rip singularity are stable to perturbation. In addition, the asymptotically static universe is stable to perturbations of this type in the static direction.
	\end{Result}
	For each of the unstable solutions we use the method of matched asymptotics (see \cite{BoundaryLayer} and the references therein) to derive formulas that approximate the (proper) time at which the singularities occur. Finally, we then \emph{numerically} investigate the stability properties of these solutions under spatially inhomogeneous mean curvature perturbations. Our findings can be summarised as follows:
	\begin{Result}
		For spatially inhomogeneous mean curvature perturbations, the oscillatory solution is unstable, with the spatial volume displaying unbounded growth. Similarly, the asymptotically static solution is unstable to mean curvature perturbations in the static direction, with the spatial volume, scalar field, and mean curvature all displaying unbounded growth. Conversely, all solutions possessing a Big Rip singularity are stable to sufficiently small mean curvature perturbations.
	\end{Result}
	It should be understood that, in the spatially inhomogeneous setting, all of our results are purely numerical. Thus, while they do indicate that solutions with a Big Rip singularity are stable, they do  not constitute a genuine proof of stability. It should be noted, however, that these results are consistent with known results for the stability of Big Rip solutions \cite{Barrow:2009,Frampton:2012}. Thus, although our results are not genuine stability proofs, we nevertheless expect them to be indicative of genuine stability.
	
	 A tabulated summary of our findings can be found in Appendix~\ref{Sec:Summary_of_our_findings}.
	\paragraph{Paper outline.} In \Sectionref{Sec:Preliminaries} we begin by recounting some basic facts about the Einstein equations and provide a brief description of near-minimal scalar field couplings. In addition, we discuss the $(3+1)$-decomposition which provides the key definitions used throughout this work. Our exact solutions are presented in \Sectionref{Sec:ExactSols}, where we provide a detailed discussion of their cosmological properties. The behaviour of these solutions under perturbation is then discussed in \Sectionref{Sec:Perturbations_and_stability}. Finally, a summary of our results is provided in \Sectionref{Sec:Conclusion}.

	\section{Preliminaries}
	\label{Sec:Preliminaries}
	The purpose of this section is to discuss the specific equations that we shall be studying throughout this work. In \Sectionref{Sec:The_equations} we discuss the Einstein equations and our particular matter coupling scheme. Then, in \Sectionref{Sec:The 3+1 decomposition}, we recall some basic facts about the $(3+1)$-decomposition and use the structure it provides to express the Einstein+matter equations as a well-posed Cauchy problem. 
	
	\subsection{The equations}
	\label{Sec:The_equations}
	Consider now a \emph{spacetime} $(M,g_{\mu\nu})$ which consists of a globally hyperbolic, $4$-dimensional, smooth, Lorentzian manifold $M$, and a smooth Lorentzian metric (i.e., a metric with signature $(-,+,+,+)$) $g_{\alpha\beta}$. Then, we study solutions of the $Z4$ system in geometric units ($c=8\pi G=1$ for the speed of light $c$ and the gravitational constant $G$), 
	\begin{subequations}
		\begin{align}
			\prescript{(4)}{}{R}_{\mu\nu} + \nabla_{\mu}Z_{\nu} + \nabla_{\nu}Z_{\mu} + \sigma\left(2\tilde{t}_{\left( \mu\right.}Z_{\left.\nu \right)} - \tilde{t}^{\mu}Z_{\mu} \right) =T_{\mu\nu}-\frac{1}{2}Tg_{\mu\nu},
			\label{Eq:EFEs}
		\end{align}
		where $\prescript{(4)}{}{R}_{\mu\nu}$ is the Ricci tensor associated with $g_{\mu\nu}$, $T_{\mu\nu}$ is the energy-momentum tensor of the matter fields with $T=g^{\mu\nu}T_{\mu\nu}$, and where $\nabla_{\mu}$ is the unique Levi-Civita connection associated with $g_{\mu\nu}$. The covector $\tilde{t}_\mu$ is considered as freely specifiable as is the constant $\sigma\in\mathbb{R}$. Conversely, the covector $Z_\mu$ is regarded as an unknown. Solutions of \Eqref{Eq:EFEs} are also solutions of the EFEs if and only if we have the additional \emph{algebraic constraint}
		\begin{align}
			Z_\mu =0.
		\end{align} 
	\end{subequations} 
	Although it may seem strange to write the EFEs in this way, we note that it is not uncommon. In fact, this particular formulation turns out to be convenient for numerical investigations. The interested reader can refer to works such as \cite{Bona:2003,Bona:2004,Bona:2005} for more details.
	
	\begin{subequations}
		In regards to the matter sector we make use of the formalism presented, by us, in \cite{Ritchie:2026}. Much of the following discussion shall only summarise the key elements that were detailed there. Suppose now that there is a real-valued scalar field $\phi:M\rightarrow\mathbb{R}$ which constitutes the matter content of the universe. The evolution equation for $\phi$ is understood as freely specifiable (see discussions in \cite{Ritchie:2026}). In this work here we shall assume that $\phi$ arises as a solution of the wave equation 
		\begin{align}
			\Box_{g}\phi = f(\phi),
			\label{Eq:EoM}
		\end{align}
		where $f(\phi)$ is some freely specifiable function. We couple this scalar field to gravity through the use of a ``coupling potential'' $V(\phi)$. The energy-momentum tensor then takes the form
		\begin{align}
			T_{\mu\nu} = \T_{\mu\nu} - V(\phi)g_{\mu\nu},
		\end{align}
		where $\T_{\mu\nu}$ is some symmetric ($\T_{\mu\nu}=\T_{\nu\mu}$) tensor which is determined as a solution of the Bianchi identity
		\begin{align}
			\nabla_{\mu}{\T^\mu}_\nu=V^{\prime}(\phi)\nabla_{\nu}\phi.
			\label{Eq:Div}
		\end{align} 	
	\end{subequations}
	We refer to $\T_{\mu\nu}$ and $V(\phi)g_{\mu\nu}$ as the ``kinetic'' and ``potential'' parts of $T_{\mu\nu}$, respectively. We therefore interpret \Eqref{Eq:Div} as an energy transfer equation. Note, however, that \Eqref{Eq:Div} is a set of $4$ equations that are intended to determine the $10$ components of $\T_{\mu\nu}$. The system is therefore underdetermined and hence some part of $\T_{\mu\nu}$ must be specified \emph{before} the system can be solved. We again refer the interested reader to \cite{Ritchie:2026} for a detailed discussion of exactly this issue.
	
	
	
	\subsection{The (3+1)-decomposition}
	\label{Sec:The 3+1 decomposition}
	
	We now recall some basic facts about the $(3+1)$-decomposition. It is in terms of this formalism that we shall choose our free data and derive exact solutions. Consider an arbitrary spacetime $(M,g_{\mu\nu})$ and suppose that there exists a smooth function $t:M\rightarrow \mathbb{R}$ whose collection of level sets $\Sigma_t$ forms a foliation $\Sigma$ of $M$. We shall assume that each $\Sigma_t\in\Sigma$ is diffeomorphic to $\mathbb{T}^3$ and hence we have 
	\begin{align}
		M = I\times\mathbb{T}^3,
		\label{Eq:Topology}
	\end{align}
	where $I$ is some time interval. This foliation yields a decomposition of $(M,g_{\alpha\beta})$ in the standard way. The unit co-normal of any $3$-surface $\Sigma_t\in\Sigma$ is 
	\begin{align}
		n_{\mu}=\alpha \nabla_{\mu}t,
		\label{eq:def_Na}
	\end{align}
	where $\alpha>0$ is the \emph{lapse}. Given the normal vector $n^\mu$ we define the induced first and second fundamental forms as
	\begin{subequations}
		\begin{align}
			\gamma_{\mu\nu}=g_{\mu\nu}+n_{\mu}n_{\nu},
			\label{Eq:Metriic_Decomp}
		\end{align}
		and
		\begin{align}
			K_{\mu\nu}=-\frac{1}{2}\mathcal{L}_{n}\gamma_{\mu\nu}.
		\end{align}
		The covariant derivative associated with $\gamma_{\mu\nu}$ is $D_{\alpha}$. The tensor field
		\begin{align}
			{\gamma^\mu}_{\nu}={\delta^\mu}_{\nu}+n^{\mu}n_{\nu},
		\end{align}
	\end{subequations}
	is the map that projects any tensor defined at any point in $M$ orthogonally to a tensor that is tangent to some $\Sigma_t$. If each index of a tensor field defined on $M$ contracts to zero with $n_{\mu}$ or $n^{\mu}$, then we call that field \textit{spatial}. Given an arbitrary tensor field on $M$ we can create a spatial tensor field on $\Sigma_t$ by contracting each index with ${\gamma^\alpha}_\beta$. In fact, any tensor can be uniquely decomposed into its spatial and its orthogonal parts, e.g.
	\begin{subequations}
		\begin{align}
			\label{eq:Tdec}
			\T_{\mu\nu}=\kappa n_{\mu}n_{\nu}+n_{\mu}j_{\nu}+n_{\nu}j_{\mu}+q_{\mu\nu},
		\end{align}   
		where
		\begin{align}
			\kappa = n^{\nu}n^{\mu}\T_{\mu\nu},
			\quad
			j_{\nu} = -{\gamma^\sigma}_{\nu}n^\mu \T_{\mu\sigma},
			\quad
			q_{\mu\nu} = {\gamma^\sigma}_{\mu}{\gamma^\eta}_{\nu}\T_{\sigma\eta}.
		\end{align}
		In our approach here we treat the field $q_{\mu\nu}$ as the ``freely specifiable'' part of $\T_{\mu\nu}$, and we additionally split it as 
		\begin{align}
			q_{\mu\nu} = \frac{1}{3}q(\kappa)\gamma_{\mu\nu} + Q_{\mu\nu},
			\quad
			Q = \gamma^{\mu\nu}Q_{\mu\nu}.
		\end{align}
		The function $q(\kappa)$ and symmetric tensor $Q_{\mu\nu}$ are constitutive freedoms that describe the properties of the scalar medium and, in particular, are freely specifiable --subject to the restriction that $Q_{\mu\nu}$ not depend explicitly on $\kappa$ or $j_\sigma$.
		
		For later use, we briefly note that the energy conditions can now be written in terms of the fields $(\kappa,Q,\phi)$. For this we first note that the isotropic pressure due to the scalar field is 
		\begin{align}
			P = \frac{1}{3}\gamma^{\mu\nu}T_{\mu\nu} = Q +q - V(\phi).
		\end{align}
		It follows then that the \emph{weak energy condition} is
		\begin{align}
			\rho\ge 0 \Longleftrightarrow \kappa+V(\phi)\ge0,
		\end{align}
		and the \emph{null energy condition} is
		\begin{align}
			\rho+P\ge0 \Longleftrightarrow \kappa + Q + q \ge 0.
			\label{Eq:Null}
		\end{align}
		One could of course also consider the strong and dominant energy conditions. However, we shall not find this necessary in the present work. 
	\end{subequations}
	
	Now pick an arbitrary vector field  $t^\mu$ such that 
	\begin{subequations}
		\begin{align}
			t^{\mu}D_{\mu}t = 1,
		\end{align}
		in fact we shall assume that $\tilde{t}_\mu$ (introduced in \Eqref{Eq:EFEs}) is chosen as $\tilde{t}_{\mu}=g_{\mu\nu}t^{\nu}$. Now, according to \Eqref{eq:def_Na} there must exist a unique spatial vector field $\beta^{\mu}$, called the \textit{shift}, such that
		\begin{align}
			t^\mu = \alpha n^{\mu} + \beta^{\mu}.
			\label{Eq:n_t}
		\end{align}
		The quantities $\alpha$ and $\beta^\mu$ are gauge freedoms that describe how a coordinate system evolves between consecutive slices $\Sigma_t$. 
	\end{subequations}
	In regards to $\nabla_{\mu}\phi$ and $Z_{\mu}$ we write	
	\begin{subequations}
		\begin{align}
			Z_{\mu} = -\Theta\, n_{\mu} + X_{\mu},
			\quad
			\nabla_{\mu}\phi = -\nu n_{\mu} + w_{\mu},
		\end{align}
		with
		\begin{align}
			\Theta = Z_{\mu}n^{\mu},
			\quad
			X_{\mu}={\gamma^\nu}_{\mu}Z_{\nu},
			\quad
			\nu = n^{\mu}\nabla_{\mu}\phi,
			\quad
			w_{\mu}={\gamma^\nu}_{\mu}\nabla_{\nu}\phi.
		\end{align}
	\end{subequations}
	Given all this, we find that $K_{\mu\nu}$ and $\gamma_{\mu\nu}$ are solutions of the evolution equations
	\begin{subequations}
		\begin{align}
			\mathcal{L}_{t}{K}_{\mu\nu}=&-D_{\mu}D_{\nu}\alpha +  \alpha( {R}_{\mu\nu} - \beta^\sigma D_{\sigma}K_{\mu\nu} +  2D_{\left(\mu\right.}X_{\left.\nu\right)} + (K+2\Theta){K}_{\mu\nu} - 2{K^\sigma}_{\mu}K_{\sigma\nu}-Q_{\mu\nu})
			\nonumber
			\\
			&+ \alpha\left(\left(\frac{q(\kappa)-3\kappa}{6} - 2\sigma\Theta\right)\gamma_{\mu\nu} +  \frac{1}{2}\left(Q - V(\phi)  \right){\gamma}_{\mu\nu} - 2K_{\mu\left(\sigma\right.} D_{\left.\nu\right)}\beta^{\sigma}\right),
			\label{Eq:Evol_K}
			\\
			\mathcal{L}_{t}\gamma_{\mu\nu}=&-2\alpha{K}_{\mu\nu} + 2\alpha D_{\left( \mu\right.}\beta_{\left.\nu \right)},
			\label{Eq:Evol_y}
		\end{align}
		where ${R^{\mu}}_\nu$ is the Ricci tensor associated with $\gamma_{\mu\nu}$, and where $K=\gamma^{\mu\nu}K_{\mu\nu}$ is the \emph{mean curvature}. Note that \Eqref{Eq:Evol_K} follows from the fully spatial projection of \Eqref{Eq:EFEs}, while \Eqref{Eq:Evol_y} is nothing more than the definition of the extrinsic curvature. In the special case $\Theta =0$ and $X_\mu =0$, then \Eqsref{Eq:Evol_K}--\eqref{Eq:Evol_y} are the standard \emph{ADM evolution equations}. In addition, the fields $\Theta$ and $X_{\mu}$ are solutions of the evolution equations
		\begin{align}
			\mathcal{L}_{t}\Theta = \alpha \beta^{\mu}D_{\mu}\Theta + \alpha H - \alpha\Theta\, K + X^{\mu}D_{\mu}\alpha + 2 \alpha D^{\mu}X_\mu - 2\sigma\alpha \Theta,
			\label{Eq:Evol_Theta}
			\\
			\mathcal{L}_{t}X_{\mu}  =\alpha M_{\mu}-2\alpha K_{\mu\nu}X^{\nu} + \Theta\, D_{\mu}\alpha - \alpha D_{\mu}\Theta - \alpha\beta^{\iota}D_{\iota}X_\mu - \alpha X_{\iota}D_{\mu}\beta^{\iota} - \sigma\alpha X_{\mu},
			\label{Eq:Evol_X}
		\end{align}
		where $R$ is the Ricci scalar and
		\begin{align}
			\label{Eq:Ham}
			H = \kappa + V(\phi) - \frac{1}{2}( R + K^{2} - K_{\mu\nu}K^{\mu\nu} ),
			\\
			\label{Eq:Mom}
			M_{\mu} = j_{\mu} + D^{\nu}K_{\mu\nu} - D_{\mu}K,
		\end{align}
		are the \emph{ADM constraint equations}. Note that \Eqsref{Eq:Evol_Theta} and \eqref{Eq:Evol_X} follow from the fully normal and mixed projections of \Eqref{Eq:EFEs}, respectively. In the special case $\Theta =0$ and $X_\mu =0$, then \Eqsref{Eq:Evol_Theta}--\eqref{Eq:Evol_X} imply $H=0$ and $M_\mu=0$. 
		
		In addition to all of this we find that \Eqsref{Eq:Div} are equivalent to the system 
		\begin{align}
			\mathcal{L}_{t}\kappa + \alpha \beta^{\mu}D_{\mu}\kappa + \alpha D^{\mu}j_{\mu}   =\, & \frac{3\kappa + q}{3}\alpha K - 2 j^\mu D_{\mu}\alpha +  \alpha Q_{\mu\nu}K^{\mu\nu}  - \alpha V^\prime(\phi)\nu,
			\label{Eq:Evol_kappa_qdd}
			\\
			\mathcal{L}_{t}j_{\sigma} + \alpha\beta^{\mu}D_{\mu}j_\sigma + \frac{1}{3}\alpha q^\prime(\kappa) D_{\sigma}\kappa =\, &  \alpha V^{\prime}(\phi)w_{\sigma} - \alpha j_{\mu}D_{\sigma}\beta^\mu - \frac{3\kappa + q}{3}D_{\sigma}\alpha + \alpha K j_{\sigma} 
			\nonumber
			\\
			&- \alpha D^{\mu}Q_{\mu\sigma} - Q_{\mu\sigma}D^{\mu}\alpha.
			\label{Eq:Evol_pd_qdd}
		\end{align}
		Finally, the equation of motion \Eqref{Eq:EoM} gives
		\begin{align}
			\mathcal{L}_{t}\phi = \alpha\nu + \alpha \beta^{\iota}w_{\iota},
			\\
			\mathcal{L}_{t}w_{\mu}  = \alpha \left( D_{\mu}\nu + \beta^{\iota}D_{\iota}w_{\mu}\right) +\alpha w_{\iota}D_{\mu}\beta^\iota +  \nu D_{\mu}\alpha, 
			\\
			\mathcal{L}_{t}\nu = \alpha ( D_{\iota}w^\iota + \beta^{\iota} D_{\iota}\nu ) + w^\iota D_\iota \alpha + \alpha \nu K - \alpha f(\phi). 
		\end{align}
		\label{Eqs:FullSystem}
	\end{subequations}
	These equations suggest the following groupings. 
	\begin{description}
		\item[Free data:] The \emph{gauge} fields $\alpha,\beta_i$ and the \emph{constitutive freedoms} $Q_{\mu\nu},q(\kappa)$ are freely specifiable everywhere on $M$ subject to the restriction that $Q_{\mu\nu}$ not depend on $\kappa$ or $j_\mu$. Similarly, the scalar field functions $V(\phi),f(\phi)$ are also freely specifiable.
		\item[Unknowns:] The fundamental forms $(\gamma_{\mu\nu},K_{\mu})$, the vectors $j_\mu,X_\mu,w_\mu$, and the scalar functions $\kappa,\phi,\nu,\Theta$ are the \emph{unknowns} which, given appropriate free data, one attempts to determine as solutions of \Eqsref{Eqs:FullSystem}.
	\end{description}
	Consider now some $t=t_0$-leaf of the foliation $\Sigma$ and suppose that the free data have been chosen. The goal is to solve \Eqsref{Eqs:FullSystem}, with initial data specified on the corresponding hypersurface $\Sigma_{t_0}$. To ensure that the resulting solutions satisfy the constraint $Z_\mu=0$ we must set
	\begin{align}
		\left. X_\mu \right|_{\Sigma_{t_0}} =0,
		\quad
		\left. \Theta \right|_{\Sigma_{t_0}}=0.
	\end{align}
	The initial data for the remaining fields must then be constructed as solutions of the constraint equations $H=0,M_\sigma=0$. It is a standard result that if the constraints are satisfied initially then the evolution equations \Eqsref{Eqs:FullSystem} ensure that they remain satisfied throughout the entire evolution.

	Suppose now that gauge freedoms $(\alpha,\beta_\mu)$ have chosen appropriately. As was discussed in \cite{Ritchie:2026}, the principal part of \Eqsref{Eq:Evol_kappa_qdd}--\eqref{Eq:Evol_pd_qdd} decouple from the remaining equations and hence their well-posedness can be discussed independently from the remaining equations. In \cite{Ritchie:2026}, we found that \Eqsref{Eq:Evol_kappa_qdd}--\eqref{Eq:Evol_pd_qdd} form a symmetric hyperbolic system with symmetrizer
	\begin{subequations}
		\begin{align}
			\frac{1}{3}\left(
			\begin{array}{cc}
				3 & 0 \\
				0 & {q}^{\prime}(\kappa){\gamma^{\sigma}}_{\nu}
			\end{array}
			\right),
		\end{align}
		provided
		\begin{align}
			{q}^{\prime}(\kappa)>0.
			\label{Eq:Hyperbolicity}
		\end{align} 
	\end{subequations}
	It can therefore be shown that, given smooth initial data for the unknowns (which has been constructed as described above), the corresponding \emph{initial value problem} is well-posed in \emph{both} the increasing \emph{and} decreasing $t$-directions. We refer to \Eqref{Eq:Hyperbolicity} as the ``hyperbolicity condition''.
	
	Observe now that all of quantities appearing in \Eqsref{Eqs:FullSystem} are smooth \emph{spatial} tensor fields. i.e., all contractions with $n^\sigma$ or $n_\sigma$ vanish. Conversely, contractions with $t^\sigma$ \emph{do not}, e.g., $X_t:=X_\sigma t^\sigma=X^\sigma t_\sigma$ as a consequence of \Eqref{Eq:n_t}. However such ``components'' $X_t$ clearly do not constitute a further degree of freedom of the field $X_\sigma$ since $X_t=X_\sigma \beta^\sigma$ is fully determined by its ``spatial components''. Consistent with this, one easily checks that the equation for $X_t$ --obtained by contracting \Eqref{Eq:Evol_X} with $t^\sigma$-- fully decouples from the remaining equations. We remark that instead of thinking of each field in the Einstein equations (\Eqsref{Eq:Evol_K}--\eqref{Eq:Evol_X}) above as a spatial field on $M$, we could equivalently think of it as a $1$-parameter family of fields on $\Sigma_t$ defined by the pull-back along the $t$-dependent map $\Lambda_t: \Sigma_t\rightarrow M$, $p\mapsto (t,p)$ to $\Sigma_t$. In the following we shall use abstract indices  $a,b,\ldots$ for such $t$-dependent tensor fields  on $\Sigma_t$. Indeed, all indices $\mu,\nu,\ldots$ in the Einstein equations above could be replaced by $a,b,\ldots$, and, at the same time, each Lie-derivative along $t^\sigma$ by the derivative with respect to parameter $t$. All this is well-known for $(3+1)$-decompositions of spacetimes and is therefore not discussed any further here.
	
	\subsection{Numerically solving the Einstein+matter equations}
	\label{Sec:Numerics}
	In this work we shall solve \Eqsref{Eqs:FullSystem} in two different ways. First, we restrict our attention to the class of Bianchi I solutions and solve the resulting equations explicitly. In a second step, we investigate spatially inhomogeneous perturbations of these exact solutions. This second step shall be done numerically. The purpose of the present subsection is to discuss how we solve \Eqsref{Eqs:FullSystem} numerically. To that end, in \Sectionref{Sec:Our_free_data_choices}, we first discuss the particular free data choices that shall be employed throughout this work. Then, in \Sectionref{Sec:NumericalSetUp}, we summarise the key components of our numerical infrastructure. A more detailed discussion of our numerical set-up, including convergence plots, can be found in Appendix~\ref{Appendix:Numerics}.  
	
	\subsubsection{Our free data choices}
	\label{Sec:Our_free_data_choices}
	Before solving \Eqsref{Eqs:FullSystem} we must first make specific choices for our free fields. To that end, suppose now we have chosen a smooth time function $t:M\rightarrow\mathbb{R}$, with the properties discussed above, giving rise to a foliation $\Sigma=\{\Sigma_t\}$ whose level sets are diffeomorphic to the torus $\mathbb{T}^3$ (see \Eqref{Eq:Topology}). We write the points in the foliation $\Sigma$ as $(t,p)$ with $t\in I$ and $p\in\mathbb T^3$. Observe carefully that we often use the same symbol $t$ for the real parameter $t\in I$ as well as for the \emph{function} $t(p)$ that defines the $(3+1)$-decomposition. Now, on each $t=$constant-slice, $\Sigma_t$, we equip the spatial manifold with coordinates $(x,y,z)\in\left(0,2\pi\right]^3$. These spatial coordinates are extended off the slices by Lie dragging along the evolution vector field $t^\mu$, satisfying $\mathcal{L}_{t}x^j=0$, for $j=1,2,3$. Consequently, $(t,x,y,z)$ forms a coordinate system that is adapted to the foliation, with the coordinate vector $\partial_{t}^\mu$ generating the flow of $t^\mu$. The gauge freedoms $(\alpha,\beta_a)$ describe how this coordinate system changes between consecutive hypersurfaces. In this work, we shall focus exclusively on \emph{geodesic slicing}, in which case we have 
	\begin{align}
		\alpha = 1,
		\quad 
		\beta_{a} = 0.
	\end{align} 
	It has been shown that for this particular gauge choice the ADM equations are only \emph{weakly} hyperbolic (see, for example, \cite{Alcubierre:Book}). We find that this is not an issue with all of our numerical solutions converging appropriately. Moving on now to the {constitutive freedoms} $(Q_{ab},q(\kappa))$, in \cite{Ritchie:2026} we found that the standard minimally coupled scalar field could be recovered from the choices 
	\begin{subequations}
		\begin{align}
			q(\kappa) = 3\kappa, 
			\quad
			Q_{ab} = w_{a}w_{b} - w_cw^c \gamma_{ab},
		\end{align}
		provided one also sets $f(\phi)=V^\prime(\phi)$.
		In the absence of any other ``physically preferable'' way to pick $(q(\kappa),Q_{ab})$ we shall pick them as above. Crucially, we note that our model corresponds to a minimally coupled scalar field only if $f(\phi)=V^\prime(\phi)$ which will not be the case here. In fact, we shall pick $f(\phi)$ as 
		\begin{align}
			f(\phi)=3\omega V(\phi),
			\label{Eq:f_choice}
		\end{align}
		for some constant $\omega\in\mathbb{R}^+$. Note here that this particular choice for $f(\phi)$ has been made for mathematical convenience only. If $f(\phi)=V^{\prime}(\phi)$ (such as occurs for a minimally coupled scalar field) then \Eqref{Eq:f_choice} implies that $V(\phi) = \mathcal{V}\e{3\omega \phi}$ for some $\mathcal{V}\in\mathbb{R}$. Given that the focus of this work is $f(\phi)\ne V^{\prime}(\phi)$ we instead choose a quadratic-type potential of the form
		\begin{align}
			V(\phi) =m_{0} + m_{1}\phi + m_2\phi^{2},
			\label{Eq:V_particular}
		\end{align}
		where $m_0,m_1,m_2\in\mathbb{R}$ are freely specifiable constants.
	\end{subequations}
	This now completely fixes our choice of free data. 
	
	In terms of these particular free data choices we find that the null energy condition, \Eqref{Eq:Null}, is
	\begin{align}
		2\kappa - w_{c}w^{c}\ge0.
		\label{Eq:Null_2}
	\end{align}
	
	\subsubsection{Outline of our numerical scheme}
	\label{Sec:NumericalSetUp}
	We now discuss how we numerically solve the Einstein+matter equations \Eqsref{Eqs:FullSystem}. The goal here is to highlight some of the key details of our numerical implementation. A more comprehensive discussion, including convergence tests, is provided in Appendix~\ref{Appendix:Numerics}. 

	To simplify our numerical procedures, we now restrict to the case for which the unknowns depend (at most) on two spatial coordinates, $x,y$. Here, any of the unknown fields defined on $M$ can be equivalently thought of as existing on a $3$-dimensional sub-manifold $\tilde{M}\subset M$ with coordinates $(t, x,y)$. In this picture, one thinks of the fields defined on $\tilde{M}$ as the pull-back to $\tilde{M}$ along the ($z$-dependent) map $\Psi_z : \tilde{M} \rightarrow M, (t, x,y)\mapsto (t, x, y, z)$. $\tilde{M}$ therefore acts as the ``effective manifold'' of our numerical implementation. Given this, \Eqref{Eq:Topology} implies
	\begin{align}
		\tilde{M} = \left[ t_-, t_+ \right]\times\mathbb{T}^2,
	\end{align}
	where some $t_-<t_+$ are constants. The specific values of $t_\pm$ are discussed when it becomes relevant. We write the points in $\tilde{M}$ as $(t,x,y)$ with $t\in\left[t_-,t_+\right]$ and $x,y\in\left(0,2\pi\right]$. All functions defined on $\tilde{M}$ can be written in terms of the standard Fourier basis. We exploit this fact in the calculation of spatial derivatives. Our code is therefore of a (pseudo-)spectral nature. To numerically implement the Fourier transform in our Python code, we use the NumPy
	Discrete Fourier Transform module\footnote{See \url{https://docs.scipy.org/doc/numpy-1.15.0/reference/routines.fft.html}} \emph{numpy.fft}. With regards to our spatial discretisation, we employ a uniform grid with $N\times N$ points.
	
	Given all of this the task now is to solve \Eqsref{Eqs:FullSystem}. To that end, in a first step, we must solve the constraint equations \Eqsref{Eq:Ham} and \eqref{Eq:Mom}. For this we use the conformal method of Lichnerowicz and York. The exact procedure is discussed in Appendix~\ref{Appendix:InitialData}. Once the initial data has been calculated we then use \Eqsref{Eqs:FullSystem} to evolve the spacetime in the \emph{increasing} $t$-direction. For our time-stepping method, we use the adaptive \emph{SciPy} integrator\footnote{See \url{https://docs.scipy.org/doc/scipy/reference/generated/scipy.integrate.odeint.html}.} \emph{odeint} with an absolute error tolerance of $10^{-12}$.
	
	In theory, if the constraints are satisfied initially then they shall remain satisfied for the entire evolution. In practice, however, numerical error means that the constraints \emph{are not} perfectly solved. 
	As a consequence it is possible for \emph{constraint violating modes} to cause run-away growth of these constraint violations. It is therefore necessary to monitor the constraint violations throughout the evolution. For any $t\in[t_-,t_+]$ we define the time-dependent quantity 
	\begin{align}
		C = \max_{x,y\in\left(0,2\pi\right]}\left( H^2 + (\delta^{-1})^{ab}M_{a}M_b \right)^{1/2},
		\label{Eq:ConstraintViolations}
	\end{align}
	which we use to monitor the constraint violations. The issue of constraint violating modes is well understood in the literature (see, for example, \cite{weyhausen2012constraint}), and can be controlled by selecting non-zero values of the constant $\sigma$ (introduced in \Eqsref{Eq:EFEs}). 


	
	\section{Exact Bianchi I solutions}
	\label{Sec:ExactSols}
	We now present our exact solutions. In this setting the precise topology of $\Sigma$ is, in some sense, irrelevant as it does not impact the final result. In fact, our particular choice of topology, \Eqref{Eq:Topology}, only becomes relevant when we consider spatially inhomogeneous perturbations	(of our exact solutions). In \Sectionref{Sec:Bianchi_I_equations}, we first derive the Bianchi I equations which are obtained by deleting all spatial covariant derivative terms appearing in \Eqsref{Eqs:FullSystem} (including the Ricci tensor $R_{ab}$). We then present and discuss our exact solutions in \Sectionref{Sec:Exact_solutions}. The stability properties of these solutions shall be discussed in \Sectionref{Sec:Perturbations_and_stability}.
	
	\subsection{The Bianchi I equations}
	\label{Sec:Bianchi_I_equations}
	We now write down the evolution equations (corresponding to \Eqsref{Eqs:FullSystem}) in the special case of Bianchi I spacetimes. In this setting, all of the unknowns (i.e., $\phi,\nu,\alpha$, and the functional components of $\gamma_{ab}$ and $K_{ab}$) depend only on the time coordinate $t$. In addition, it is convenient to split $K_{ab}$ into its trace and trace-free parts. i.e., 
	\begin{subequations}
		\label{Eq:Bianchi1Reduction}
		\begin{align}
			{K^{a}}_{b}=\frac{1}{3}K {\gamma^{a}}_{b}+{\chi^{a}}_{b},
			\quad
			\chi^a{}_{a}=0,
			\quad
			K = {K^{a}}_{a}.
		\end{align}
		In regards to the metric, we suppose that there is a scalar function $\Omega:M\rightarrow \mathbb{R}$ and a tensor $\tilde{\gamma}_{ab}$ such that 
		\begin{align}
			\gamma_{ab}=\Omega^{2}\tilde{\gamma}_{ab},
			\quad
			\tilde{\gamma}_{ab} = \text{diag}( \gamma_{1},\gamma_{2},\gamma_{3} ).
			\label{Eq:SpecialMetric}
		\end{align}
		Here we have restricted our attention to diagonal metrics. This is not a significant restriction as, for spatially homogeneous solutions, one can always perform a (local) coordinate transformation so that the metric $\gamma_{ab}$ is diagonal. 
	\end{subequations}
	
	Given all of this we find that the evolution equations for the first and second fundamental form can be written as   
	\begin{subequations}
		\label{Eqs:HomoEvolve}
		\begin{align}
			\partial_{t}\Omega = -\frac{1}{3} K \Omega,
			\quad
			\partial_{t}\tilde{\gamma}_{ab}=-2\tilde{\gamma}_{ac}{{\chi^c}_{b}},
		\end{align}
		and
		\begin{align}
			\partial_{t}K = K^2 - 3 V(\phi),
			\quad
			\partial_{t}{\chi^{a}}_{b}= K {\chi^{a}}_{b}.
		\end{align}
		The evolution equations for $\tilde{\gamma}_{ab}$ and ${\chi^a}_b$ can be further reduced by setting 
		\begin{align}
			\gamma_{i} = \tilde{\gamma}_{i}\exp\left( -2{C_i}p(t) \right),
			\quad
			\chi^a{}_b = \chi(t)\text{diag}(C_1,C_2,C_3),
		\end{align}
		where, for each $i=1,2,3$, we have that $\tilde{\gamma}_{i},C_i\in\mathbb{R}$ are integration constants, and where the functions $p(t)$ and $\chi(t)$ are solutions of the evolution equations 
		\begin{align}
			\partial_{t}p(t) = \chi(t),
			\quad
			\partial_{t}\chi(t)=K \chi(t).
			\label{Eq:p_and_u}
		\end{align}
		Note that, since ${\chi^a}_b$ is trace-free, we must have 
		\begin{align}
			C_1 + C_2 + C_3 = 0.
			\label{Eq:C_i}
		\end{align}
	\end{subequations}
		Universes with metrics of this type \Eqref{Eq:SpecialMetric}, are understood to be expanding (or contracting) at different rates in each of the three spatial directions, with isotropy occurring only in the special case $C_{1}=C_{2}=C_{3}=0$. We define the directional scale factors $a_i(t)$ in the standard way: 
		\begin{subequations}
			\begin{align}
				a_{i}(t)=\Omega(t)\sqrt{\gamma_{i}(t)}.
				\quad
				i=1,2,3.
			\end{align}
			The ``velocity'' of spatial expansion (or contraction) in a given direction is captured by the directional Hubble parameters $H_i(t)$, which are defined as 
			\begin{align}
				H_{i}(t)=\frac{\partial_{t}a_{i}(t)}{a_{i}(t)}=-\frac{1}{3}K(t) - C_{i}\chi(t),
				\label{Eq:Hubble}
			\end{align}
			while the \emph{average scale factor} $a_{av}(t)$ can be calculated as 
			\begin{align}
				a_{av}(t)=(a_{1}(t)a_{2}(t)a_{3}(t))^{1/3} = \Omega(t),
			\end{align}
			where we have used that, as a consequence of \Eqref{Eq:C_i}, ${\gamma}_{1}{\gamma}_{2}{\gamma}_{3}=1$. From here we can see that $\Omega$ represents the isotropic scale of the universe while $\Omega^3$ describes its volume. 
		\end{subequations}
		
		Turning our attention to the scalar sector, we find that the evolution equations for $\nu$ and $\phi$ are 
		\begin{subequations}
		\begin{align}
			\partial_{t}\nu = K \nu - 3\omega V(\phi),
			\quad
			\partial_{t}\phi =  \nu.
			\label{Eq:nu_phi_eqs_Bianchi1}
		\end{align}
		The remaining evolution equation is 
		\begin{align}
			\partial_{t}\kappa = 2\kappa K - \partial_{t}V(\phi),
		\end{align}
		and the Hamiltonian constraint is 
		\begin{align}
			\kappa = \frac{1}{2}\left( K^{2} - K_{ab}K^{ab} \right) - V(\phi)=\frac{1}{3}K^{2}-\frac{1}{2}{\chi^a}_{b}{\chi^b}_{a}-V(\phi). 
			\label{Eq:Hamiltonian_Bianchi_1}
		\end{align}
	\end{subequations}
	One can check via direct calculations that $\kappa$, as given by the Hamiltonian constraint, is always a solution of the evolution equation for $\kappa$ and hence $\kappa$ can be calculated \emph{after} $K$ and $\phi$ have been determined. In addition, we note that (in this spatially homogeneous setting) the null energy condition \Eqref{Eq:Null_2} reduces to $\kappa\ge0$.

	It turns out that, as a direct consequence of \Eqref{Eq:f_choice}, the evolution equation for $\nu$ can be solved exactly as
	\begin{subequations}
		\begin{align}
			\nu(t)=\omega K(t) + \nu_{\star}\chi(t),
			\label{Eq:nu_particular}
		\end{align}
		where $\nu_{\star}\in\mathbb{R}$ is an integration constant. We have now parametrised both $\chi^a{}_{b}$ and $\nu$ in terms of the function $\chi(t)$. In order to ensure that \Eqref{Eq:p_and_u} does not generate trivial solutions for $\chi$ it is necessary to assume that there is at least one time $t_\star$ such that $\chi(t_\star)\neq0$. If $\chi$ does not contribute the geometry (or to $\nu$) then this can be controlled by setting $C_i=0$ (or $\nu_\star=0$).		
	\end{subequations}

	\subsection{Exact solutions}
	\label{Sec:Exact_solutions}
	Let us now construct explicit solutions of the Bianchi I equations, presented above. For this, we restrict our attention to the special case $\nu_{\star}=0$ (see \Eqref{Eq:nu_particular}) so that $\nu=\omega K$. In this setting, we find that the exact solutions for $\Omega$ and $\chi$ can be expressed in terms of $\phi$ as\footnote{In some sense the scalar field here is being used to parametrise time. Typically, when doing this one sets $t=t(\phi)$ and assumes that $\phi$ is strictly monotonic (so that the equation is invertible). We have \emph{not} assumed this. Instead, we note that \Eqsref{Eq:Omega} define conservation relationships between $\Omega,\chi$ and $\phi$ which can then be used to determine $\Omega$ and $\chi$ \emph{algebraically} in terms of $\phi$. The validity of these relations can be verified by use of the evolution equations \emph{without} the assumption of strictly monotonic $\phi$.}
	\begin{subequations}
		\begin{align}
			\Omega(t) = \Omega_{0}\exp\left( -\frac{1}{3\omega}\phi(t) \right),
			\quad
			\chi(t) = \chi_0\exp\left(\frac{1}{\omega}\phi(t)\right),
			\label{Eq:Omega}
		\end{align}
		where $\Omega_{0},\chi_0\in\mathbb{R}$ are integration constants. Note that one can always rescale the spatial coordinates by a factor of $\Omega_{0}$ and hence $\Omega_{0}$ can be viewed as a coordinate freedom. Thus, without loss of generality, we set $\Omega_{0}=1$.	Similarly, we set $\chi_0=1$.
	\end{subequations}

	Given this we find that the mean curvature $K$ can be written as 
	\begin{subequations}
		\begin{align}
			\label{Eq:K_of_phi}
			\begin{split}
				K^{2} =&\, 3m_2\left(\phi - \phi_\star \right)^2  -\frac{3m_\star}{4m_2} +\frac{3}{2} \mathcal{K}_\star \exp\left( \frac{2}{\omega}\phi \right),
			\end{split}
		\end{align}
		where $\mathcal{K}_\star\in\mathbb{R}$ is an integration constant and  
		\begin{align}
			m_{\star} = m_1^2 - 4m_0 m_2 - m_2^2\omega^2,
			\quad
			\phi_{\star} = -\frac{m_1+\omega m_{2}}{2m_{2}}. 
		\end{align}
	\end{subequations}
	
	Turning our attention now to $\kappa$, \Eqref{Eq:Hamiltonian_Bianchi_1} implies
	\begin{align}
		\kappa=  \frac12\omega (m_1 + m_2\omega) + (m_1 + m_2\omega)\phi  - \frac{1}{2}\e{2\phi/\omega}(C_1^2+C_2^2+C_3^2-\mathcal{K}_\star).
		\label{Eq:kappa}
	\end{align}
	While $\mathcal{K}_\star$ is in general non-zero, in this subsection here we shall restrict our attention to the choice $\mathcal{K}_\star=0$ as it ensures that the resulting equation for $\phi$ is integrable. In fact, we find that 
	\begin{subequations}
		\label{Eq:Bianchi1_remaining}
		\begin{align}
			(\partial_{t}\phi)^2 = 3m_2w^{2}\left(\phi - \phi_\star \right)^2  -\frac{3w^{2}m_\star}{4m_2}.
			\label{Eq:phi_quad}
		\end{align}
		Note that, in \Eqref{Eq:phi_quad}, it is necessary to make a specific choice of the sign of $\partial_t\phi$. The impact of this choice is discussed below. In addition, the evolution equation for $p(t)$ is  
		\begin{align}
			\partial_{t}p(t)={\exp(\phi(t)/\omega)}.
			\label{Eq:p_of_phi}
		\end{align}
	\end{subequations}
	It now only remains to solve \Eqsref{Eq:Bianchi1_remaining}. Both \Eqsref{Eq:phi_quad} and \eqref{Eq:p_of_phi} are integrable. However, the precise structure of the solution depends on the sign of $m_\star$ and $m_2$. In total we find that there are four distinct cases: $(m_2<0,m_\star>0),(m_2>0,m_\star>0),(m_2>0,m_\star<0)$, and $(m_2>0,m_\star=0)$. If $m_2<0$ and $m_\star\le 0$ then we find that $K^{2}\le 0$ and as such this case has been excluded. For each case we present the exact solution and discuss the properties of the corresponding cosmology. For this we find it is useful to define the following constants 
	\begin{align}
		\lambda = \sqrt{\frac{|m_\star|}{4 m_2^2}},
		\quad
		\omega_{\star}=\sqrt{3|m_2|}\omega,
		\quad
		p_{\star} = \e{\phi_\star/\omega},
		\quad
		\kappa_{\star} = \frac{\omega}{2}\left(  m_0 + m_1\omega - 4m_{2}\phi_{\star}^2  \right).
	\end{align}
	We now discuss each of the different solutions. For our discussion here we restrict our attention to $\omega>0$. Although not considered here, we emphasise that analogous results can be derived for $\omega<0$.
	
	We remind the reader that, a tabulated summary of our solutions and their asymptotic dynamics, can be found in Appendix~\ref{Sec:Summary_of_our_findings}. 
	\subsubsection{Case 1: An oscillatory universe}
	\label{Sec:Case_1}
	We first consider the case $(m_2<0,m_\star>0)$. For this particular parameter regime, solutions of \Eqref{Eq:phi_quad} take the form\footnote{Recall here that, according to \Eqref{Eq:phi_quad}, {the sign of $\partial_{t}\phi$ is freely specifiable. This is precisely the origin of the $\pm$ in} \Eqref{Eq:phi_osc}.}
	\begin{align}
		\phi(t) = \phi_{\star}  \pm \lambda\cos\left( \omega_{\star}(t-t_\star) \right),
		\label{Eq:phi_osc}
	\end{align} 
	where $t_\star\in\mathbb{R}$ is an integration constant. The $\pm$ can be chosen freely. For our discussion here we shall take the positive sign although it should be noted that similar results can be derived for the case of a minus sign. From \Eqref{Eq:phi_osc}, it is clear that the scalar field, and hence the averaged scale factor $\Omega$, is oscillatory with period $\Delta t=2\pi/\omega_\star$. We can exploit this fact to integrate \Eqref{Eq:p_of_phi}, by first noting that\footnote{This expansion can be found in \cite{MathematicalFunctions}, Eq. ({9.6.34}).}
	\begin{subequations}
		\begin{align}
			\e{\lambda \cos( \omega_\star(t-t_\star))/\omega}   = I_{0}(\lambda/\omega) + 2\sum_{n=1}^{\infty}I_{n}(\lambda/\omega)\cos(n\,\omega_{\star}(t-t_\star)),
			\label{Eq:JacobiAngler_Case_1}
		\end{align}
		where, for each fixed $n=0,\dots,\infty$, we have 
		\begin{align}
			I_{n}(z) =\sum_{k=0}^{\infty}\frac{1}{k!(n+k)!}\left( \frac{z}{2} \right)^{2k+n},
			\label{Eq:ModifeidBesselDef}
		\end{align}
	\end{subequations}
	is the $n$th modified Bessel function of the first kind. Plugging this expansion into \Eqref{Eq:p_of_phi} and integrating the result reveals 
	\begin{align}
		p(t) = p_{\star}I_{0}\left(\frac{\lambda}{\omega}\right)t + \underbrace{	2p_{\star}\sum^{\infty}_{n=1}\frac{I_{n}(\lambda/\omega)}{n\, \omega_{\star}}\sin(n\,\omega_{\star}(t-t_\star))}_{:=\hat{p}(t)}
		= p_{\star}I_{0}\left(\frac{\lambda}{\omega}\right)t + \hat{p}(t).
	\end{align}
	Putting all of this together we find that the spacetime line-element is
	\begin{subequations}
		\begin{align}
			\begin{split}
				ds^2 =& -dt^2 + \e{-\frac{2\lambda}{3\omega}\cos(\omega_{\star}(t-t_0))}dR^2(I_{0}({\lambda}/{\omega}),\hat{p}),
			\end{split}
		\end{align}
		where we have defined the ($h,\hat{p}$-dependent) spatial line-element $dR^2(h,\hat{p})$ as 
		\begin{align}
			\begin{split}
				dR^2(h,\hat{p})= \e{-2C_1p_\star h} \e{-2C_1\hat{p}(t)}dx^2 +&  \e{-2C_2p_\star h} \e{-2C_2\hat{p}(t)}dy^2 
				\\
				&+ \e{-2C_3p_\star h} \e{-2C_3\hat{p}(t)}dz^2.
			\end{split}
			\label{Eq:dR2}
		\end{align}
	\end{subequations}
	We now discuss the key features of this cosmology. As stated above, this universe is oscillatory. Moreover, it is possible to show that the curvature invariants (such Kretschmann scalar) remain finite for all $t\in(-\infty,\infty)$. We therefore conclude that this solution is both past and future geodesically complete. i.e., there are no singularities.
	
	We now discuss the behaviour if this universe during a single period of the scalar field's oscillation. For this, we first define 
	\begin{subequations}
		\begin{align}
			\phi_{\pm}=\phi_{\star} \pm \lambda,
		\end{align}
		we find that $\phi\in[\phi_-,\phi_+]$. As for the potential, $V(\phi)$, we get 
		\begin{align}
			V(\phi_\pm) = \pm {m_2\omega}\lambda. 
		\end{align}
		It follows then that the potential changes sign as the scalar field $\phi$ oscillates. In addition to all of this we can define
	\end{subequations}
	\begin{subequations}
		\begin{align}
			\kappa_{\pm} = \kappa_\star \pm (m_1+m_2\omega)\lambda - \frac{1}{2}\exp\left( -\frac{m_1+\omega m_2}{m_2\omega} \pm \frac{2\lambda}{\omega}\right)(C_1^2+C_2^2+C_3^2),
		\end{align}
		and hence we have 
		\begin{align}
			\min\{ \kappa_{-},\kappa_{+} \}\le\kappa\le \max\{ \kappa_{-},\kappa_{+} \}.
		\end{align}
		There are two possibilities here, depending on the size of the constants $C_i$. On the one hand, it is possible that the null energy condition is never satisfied. On the other hand, if $C_i\ll1$, and 
		\begin{align}
			\kappa_\star >-| (m_1+m_2\omega)\lambda|,
		\end{align} 
		then it is possible to ensure that the null energy condition, \Eqref{Eq:Null_2}, is always satisfied.
	\end{subequations}
	
	We now turn our attention now to the (average) scale factor $\Omega(t)$, and make use of \Eqsref{Eqs:HomoEvolve} to find that 
	\begin{align}
		\Omega^{-1}{\partial_{t}^{2}\Omega} =\frac{1}{9}\left( 9V(\phi) - 2K^{2} \right). 
	\end{align}
	Then, following from \Eqref{Eq:phi_osc}, we have 
	\begin{subequations}
		\begin{align}
			({\partial_{t}^{2}\Omega})(t_c)=0\implies t_c=\arccos\left( \frac{3\omega - \sqrt{9\omega^2+4\lambda^2}}{2\lambda} \right).
		\end{align}	
		This now implies that we can divide each cycle into four phases: Each cycle starts at 
		\begin{align}
			t_{0}=t_{\star} + \frac{2n\pi}{\omega_{\star}},
			\label{Eq:StartTime}
		\end{align}
		for some integer $n$. At $t=t_0$ the scalar field is at its largest value while the volume element ($\Omega^3$) is at its smallest. Moreover, at this time, we find that $K(t_0)=0$. There are four more times of interest: 
		\begin{align}
			\begin{split}
				\label{Eq:times}
				t_{1}=t_0 + \frac{t_c}{\omega_{\star}},
				\quad
				t_{2}=t_{0} + \frac{\pi}{\omega_\star},
				\quad
				t_{3}=t_{0}+\frac{2\pi-t_c}{\omega_{\star}},
				\quad
				t_{4}=t_{0}+\frac{2\pi}{\omega_{\star}}.
			\end{split}
		\end{align}
	\end{subequations}
	For $t\in(t_0,t_2)$ the volume of the universe increases until it reaches its largest values at $t=t_2$. This expansion is accelerated for $t\in(t_0,t_1)$ and decelerated for $t\in(t_1,t_2)$. For $t\in(t_2,t_4)$ the volume decreases until it once again reaches its minimum value at $t=t_4$. This contraction is decelerated for $t\in(t_2,t_3)$ and accelerated for $t\in(t_3,t_4)$. A summary of these four phases is provided in Table~\ref{Table:PhasesOscillate}.
	\begin{table}[t!]
		\centering
		\begin{tabular}{|c|c|c|c|c|}
			\hline
			Phase & Interval & $\partial_{t}\Omega$ & $\partial_{t}^{2}\Omega$ & Physical state  \\
			\hline
			I: Recovery & $(t_{0},t_1)$ & $>0$ & $>0$ & Accelerated expansion \\
			\hline
			II: Coasting & $(t_1,t_2)$ & $>0$ & $<0$ & Decelerated expansion \\
			\hline
			III: Fall & $(t_{2},t_3)$ & $<0$ & $<0$  & Decelerated contraction \\
			\hline
			IV: Cushion & $(t_{3},t_{4})$ & $<0$ & $>0$ & Accelerated contraction \\
			\hline
		\end{tabular}
		\caption{\label{Table:PhasesOscillate}Summary of the various phases of the universe during a single period of the scalar field oscillation. The starting time $t_0$ is defined in \Eqref{Eq:StartTime} and the remaining times, $t_0,t_1,t_2,t_3,t_4$, are given in \Eqref{Eq:times}.}
	\end{table}
	In the special case of isotropy (defined by $C_1=C_2=C_3=0$) the oscillatory behaviour described above forms a complete picture of the universes dynamics. However, if the universe is anisotropic the story is slightly more complicated. Turning our attention to the Hubble parameters $H_i$ (see \Eqref{Eq:Hubble}), we find
	\begin{align}
		H_{i}=-\frac{1}{3}K - C_{i}\e{\phi(t)/\omega}.
	\end{align}
	This expression reveals a competition between two distinct physical effects: The $-K/3$-term represents the symmetric ``breathing'' of the universe, which averages to zero over a period $\Delta t$, while the $C_{i}\e{\phi(t)/\omega}$-terms represent the anisotropic drift. To find the net physical effect over time, we integrate the expansion rate over one cycle. The resulting averaged Hubble parameter, $\left<H_{i}\right>$, determines the long-term spectral shift:
	\begin{align}
		\left<H_{i}\right> = -C_{i}\e{-\frac{m_1+\omega m_2}{2\omega m_2}}I_{0}\left( \frac{\lambda}{\omega} \right).
	\end{align}
	Let $z_i$ be the \emph{spectral shift}\footnote{Note that spectral shift $z_i$ is calculated via the standard ``redshift'' formula. However, $z_i$ itself can only be interpreted as redshift if $z_i>0$. If $z_i<0$ then $z_i$ describes blueshift instead (see, for example, \cite{Fleury:2015}). Given that the second case is most relevant for the present work, we refer to $z_i$ as the ``spectral shift''.} in the $x^i$-direction. Then, 
	\begin{align}
		1 + \left<z_i\right> \approx \e{\left<H_i\right>\Delta t}
	\end{align}
	It follows then that the constants $C_i$ impact the measured spectral shift. Along any axis where $C_i>0$ the average Hubble parameter is negative (i.e., $\left<H_{i}\right><0$). Physically, this implies that a photon emitted from a distant source will be compressed by the net contraction of that specific dimension. An observer would therefore measure a net blueshift in directions with $C_i>0$ and a redshift in directions with $C_i<0$. To understand why this happens we note that, following from \Eqref{Eq:C_i}, at least one of the constants $C_i$ must be positive. Thus, even though the net change in the volume (of the universe) over a single cycle vanishes, individual directions are contracting (or expanding). This implies the existence of an effective ``optical-horizon'': Light travelling in the direction of contraction would be continuously blueshifted, potentially reaching extreme energies, while light travelling with the drift would be redshifted, leading to exponential suppression of the observed signal, making it observationally negligible
	
	\subsubsection{Case 2: Bracketed by singularities}
	\label{Sec:Case_2}
	We now consider the case $(m_2>0,m_\star>0)$. In contrast to the oscillatory behaviour seen in Case 1, the solutions for $\phi$ (arising as a consequence of \Eqref{Eq:phi_quad}) take the hyperbolic form
	\begin{align}
		\phi = \phi_\star \pm \lambda\cosh\left( \omega_{\star}(t-t_\star) \right),
		\label{Eq:phi_2}
	\end{align} 
	where $t_\star\in\mathbb{R}$ is an integration constant. The scalar field, as given in \Eqref{Eq:phi_2}, describes a universe that is singular in both the decreasing \emph{and} increasing $t$ directions. While Case 1 described a universe that ``breathes'' indefinitely, Case 2 describes a cosmology that exists for an infinite period of proper time between two singular endpoints. 
	
	Before discussing this further, let us first integrate \Eqref{Eq:p_of_phi} and construct the function $p(t)$. To this end, we note that\footnote{A derivation of this formula can be found in the supplementary material.}
		\begin{align}
			\e{\pm\lambda \cosh( \omega_\star(t-t_\star))/\omega}   = I_{0}(\lambda/\omega) + 2\sum_{n=1}^{\infty}(\pm1)^{n}I_{n}(\lambda/\omega)\cosh(n\,\omega_{\star}(t-t_\star)),
		\end{align}
		where we have once again made use of \Eqref{Eq:ModifeidBesselDef}. Plugging this expansion into \Eqref{Eq:p_of_phi} and integrating the result reveals 
	\begin{align}
		p(t) = p_\star I_{0}\left(\frac{\lambda}{\omega}\right)t +\underbrace{2p_\star \sum^{\infty}_{n=1}(\pm1)^{n}\frac{I_{n}(\lambda/\omega)}{n\, \omega_{\star}}\sinh(n\,\omega_{\star}(t-t_\star))}_{:=\hat{p}(t)} = p_\star I_{0}\left(\frac{\lambda}{\omega}\right)t + \hat{p}(t).
		\label{Eq:phat_case_2}
	\end{align}
	Putting all of this together we find that the spacetime line-element is 
	\begin{align}
		\begin{split}
			ds^2 =& -dt^2 + \e{\mp\frac{2\lambda}{3\omega}\cosh(\omega_{\star}(t-t_0))}dR^2(I_{0}({\lambda}/{\omega}),\hat{p}),
		\end{split}
	\end{align}
	where $dR^2(h,\hat{p})$ is defined as in \Eqref{Eq:dR2} with $\hat{p}$ given as in \Eqref{Eq:phat_case_2}. In order to analyse the dynamics of this universe, we calculate the (leading order of the) Kretschmann scalar\footnote{We use the $O$ symbol rather informally in the usual sense $f = O(g) \iff |f| \le C |g| $ for some constant $C>0$ in the relevant limit.}, $\mathcal{K}$:
	\begin{align}
		\mathcal{K} = \left(8m_2\, \phi^4 + O\left( \phi^3 \right)\right) + O\left( \phi^4 \e{\frac{3}{\omega}\phi} \right).
		\label{Eq:Kr}
	\end{align}
	Thus, the Kretschmann scalar becomes infinite if the scalar field does. Note that this is true for all of the exact solutions presented here. It follows then that this universe has two physical singularities, which occur in the limits $t\rightarrow\pm \infty$. However, the nature of these singularities changes, depending on the chosen sign (in \Eqref{Eq:phi_2}). On the one hand, if the positive sign is chosen then $\Omega\rightarrow0$ in the limit $|t|\rightarrow\infty$. This case therefore corresponds to a crushing singularity. i.e., the universe starts with a Big Bang, expands until it reaches its maximum volume at $t=t_\star$, contracts and ends in a Big Crunch. On the other hand, if a minus sign is chosen then $\Omega\rightarrow\infty$ in the limit $|t|\rightarrow\infty$. This is a ``Big Rip'' singularity\footnote{It should be noted that this singularity occurs in infinite proper time and as such can never truly be reached by an observer. Big Rip type singularities that are formally unreachable are sometimes also called ``Little Rip''. See, for example, \cite{Frampton:2012}.} with the universe obtaining its minimum volume at $t=t_\star$.
	\begin{subequations}
		In regards to $\kappa$, we find that 
		\begin{align}
			\begin{split}
				\kappa = &\underbrace{\kappa_\star \pm (m_1+m_2\omega)\lambda\cosh(\omega_{\star}(t-t_0))}_{I} 
				\\
				&\quad - \frac{1}{2}\underbrace{\e{-\frac{m_1+\omega m_2}{2\omega m_2}}\e{\pm \frac{\lambda}{\omega}\cosh\left( \omega_{\star}(t-t_\star) \right)}(C_1^2+C_2^2+C_3^2)}_{II}.
			\end{split}
		\end{align}
		The sign here becomes quite important. Regardless of the chosen sign, one can always chose $m_1$ in such a way as to ensure that term I is positive near the singularities (and hence the null energy condition \Eqref{Eq:Null_2} is satisfied). However, the anisotropy (represented in term II) always contributes negatively to $\kappa$. If the positive sign is chosen (and there is at least one $C_i\ne0$) then term II becomes infinite near $t=\pm\infty$ and hence the null energy condition is always violated. Conversely, if the minus sign is chosen, then term II approaches zero in the limit $|t|\rightarrow\infty$. In this case we find that, if
		\begin{align}
			\begin{split}
				{\kappa_\star - (m_1+m_2\omega)\lambda} - \frac{1}{2}{\e{-\frac{m_1+\omega m_2}{2\omega m_2}}(C_1^2+C_2^2+C_3^2)}\ge0
			\end{split}
		\end{align}
		then the null energy condition is satisfied for all $t\in(-\infty,\infty)$. This identifies the negative branch as a physically viable, albeit divergent, cosmology
	\end{subequations}
	
	\subsubsection{Case 3: A Big Bang and a Big Rip}
	\label{Sec:Case_3} 
	We now consider the case $(m_2>0, m_\star<0)$. For this parameter regime, the solutions to \Eqref{Eq:phi_quad} take the form:
	\begin{align}
		\phi(t) = \phi_\star \pm \lambda\sinh\left( \omega_{\star}(t-t_\star) \right),
	\end{align}
	where $t_\star\in\mathbb{R}$ is an integration constant. To integrate the metric function $p(t)$ via \Eqref{Eq:p_of_phi}, we utilize the Jacobi-Anger-like expansion for the modified Bessel functions\footnote{A derivation of this formula can be found in the supplementary material.}:
	\begin{align}
		\e{\pm z \sinh \theta} = J_0(z) + 2 \sum_{n=1}^{\infty} J_{2n}(z) \cosh(2n\theta) \pm 2 \sum_{n=0}^{\infty} J_{2n+1}(z) \sinh((2n+1)\theta),
	\end{align}
	where $J_{n}(z)$ are the Bessel functions of the first kind. Substituting $z = \lambda/\omega$ and $\theta = \omega_\star(t-t_\star)$, we integrate term-by-term to find that $p(t)$ has the structure
	\begin{subequations}
		\begin{align}
			p(t) = p_\star J_0\left(\frac{\lambda}{\omega}\right)t + \hat{p}(t),
		\end{align}
		where 
		\begin{align}
			\label{Eq:phat_case_3}
			\begin{split}
				\hat{p}(t) =  \frac{2p_\star}{\omega_{\star}} \sum_{n=1}^{\infty} \frac{J_{2n}({\lambda}/{\omega})}{2n} \sinh(2n\theta) \pm \frac{2p_\star}{\omega_{\star}} \sum_{n=0}^{\infty} \frac{J_{2n+1}({\lambda}/{\omega})}{2n+1} \cosh((2n+1)\theta),
			\end{split}
		\end{align}
		and where, for the sake of readability, we have used $\theta=\omega_\star(t-t_\star)$.
	\end{subequations}
	Putting all of this together we find that the spacetime line-element is 
	\begin{align}
		\begin{split}
			ds^2 =& -dt^2 + \e{\mp\frac{2\lambda}{3\omega}\sinh(\omega_{\star}(t-t_0))}dR^2(J_{0}({\lambda}/{\omega}),\hat{p}),
		\end{split}
	\end{align}
	where $dR^2(h,\hat{p})$ is defined as in \Eqref{Eq:dR2} and $\hat{p}$ is given by \Eqref{Eq:phat_case_3}.	In order to analyse the dynamics of this universe, proceed as in the previous case. i.e., we calculate the Kretschmann scalar, $\mathcal{K}$. In doing so, we once again find that $\mathcal{K}=O(\phi)$. Thus, we again conclude that this universe has two physical singularities, which occur in the limit $|t|\rightarrow \infty$. \emph{Unlike} our previous case, the type of singularity is not the same in either direction. If a minus sign is chosen, then the universe begins in a Big Bang (at $t=-\infty$) and ends in a Big Rip at $t=+\infty$. If a positive sign is chosen the dynamics is reversed. 
	
	In regards to $\kappa$, we find that 
	\begin{align}
		\begin{split}
			\kappa = &{\kappa_\star \pm (m_1+m_2\omega)\lambda\sinh(\omega_{\star}(t-t_0))}
			\\
			&\quad - \frac{1}{2}{\e{-\frac{m_1+\omega m_2}{2\omega m_2}}\e{\pm \frac{\lambda}{\omega}\sinh\left( \omega_{\star}(t-t_\star) \right)}(C_1^2+C_2^2+C_3^2)}.
		\end{split}
	\end{align}
	In this case, regardless of sign, it is possible to arrange that the null energy condition \Eqref{Eq:Null_2} is satisfied near one of the singularities only. For example, in the case of a positive sign we find that if $m_1+m_2\omega<0$ then $\kappa\rightarrow\infty$ as $t\rightarrow-\infty$ and $\kappa\rightarrow-\infty$ as $t\rightarrow\infty$.

	\subsubsection{Case 4: Singular and static} 
	\label{Sec:Case_4}
	For our last case, we consider the parameter regime $(m_2>0,m_\star=0)$. In this setting, solutions of \Eqref{Eq:phi_quad} take the form
	\begin{align}
		\phi = \phi_\star + \exp\left( \pm\omega_{\star}(t-t_\star) \right),
	\end{align}
	where $t_{\star}\in\mathbb{R}$ is an integration constant. In this case we restrict our attention to the minus sign. Analogous results can be derived in the case of a positive sign. Integrating \Eqref{Eq:p_of_phi} reveals that $p(t)$ takes the form
	\begin{subequations}
		\begin{align}
			p(t) = -\frac{1}{\omega_{\star}}\e{-\frac{m_1+\omega m_2}{2m_2} }\mathrm{Ei}\left(\e{-\omega_{\star}(t-t_0)}\right),
		\end{align}
		where
		\begin{align}
			\mathrm{Ei}\left( z \right)
			=
			-\operatorname{PV}\int_{-z}^{\infty}\frac{\text{e}^{-s}}{s}\,ds\implies \frac{d\mathrm{Ei}(z)}{dz}=\frac{\e{z}}{z},
		\end{align}
	\end{subequations}
	is the exponential integral function. Here the principal value prescription is understood to be due to the logarithmic singularity of the integrand at $s=0$. In particular, $Ei(z)\sim \gamma+\ln(z)$ as $z\rightarrow0^+$, which determines the asymptotic behaviour of $p(t)$. Note that $\gamma$ is the Euler–Mascheroni constant.
	
	Putting all of this together we find that the spacetime line-element is 
	\begin{align}
		\begin{split}
			ds^2 = -dt^2 + \e{2\e{-\omega_{\star}(t-t_\star)}}\left( \e{-2C_1{p}(t)}dx^2  +  \e{-2C_2{p}(t)}dy^2 + \e{-2C_3{p}(t)}dz^2\right).
		\end{split}
	\end{align} 
	Calculating the Kretschmann scalar\footnote{Here we have made use of the shorthand notation $O(g,h)$ to mean that, if $f = O(g,h) \iff |f| \le C (|g|+|h|) $ for some constant $C>0$ in the relevant limit.} 
	\begin{align}
		\mathcal{K}= O\left( \e{-2\omega_{\star}(t-t_\star)}, C_{i}\e{ 2\e{-\omega_{\star}(t-t_\star)}/\omega } \right).
	\end{align}
	Here we see that, if $t\rightarrow-\infty$ then $\mathcal{K}\rightarrow\infty$ and hence $t=-\infty$ is a physical singularity. In this same limit we find that $\Omega\rightarrow0$. Conversely, we find that $\mathcal{K}$ remains finite in the limit $t\rightarrow\infty$. Moreover, we find that $\Omega\rightarrow\e{\frac{m_1+m_2\omega}{3\omega}}$ as $t\rightarrow\infty$. This solution therefore represents a universe that has a Big Bang type singularity in the infinite past and approaches a static universe in the infinite future. 
	
	Turning our attention to $\kappa$, we find that 
	\begin{align}
		\kappa = \kappa_{\star} + (m_1+m_2\omega)\omega\e{-\omega_{\star}(t-t_\star)} - \frac{1}{2}\e{\frac{m_1+\omega m_2}{\omega m_2}}(C_1^2+C_2^2+C_3^2)\e{\frac{2}{\omega}\e{-\omega_{\star}(t-t_\star)}}.
	\end{align} 
	In the limit $t\rightarrow-\infty$ we find that $\kappa>0$ only if and only if $m_1+m_2\omega>0$ \emph{and} $C_1=C_2=C_3=0$. If the solution is anisotropic then the super exponential ensures that $\kappa<0$ for all $t$ near $t=-\infty$. Conversely, if $t\rightarrow\infty$, we find that the null energy condition is satisfied if the inequality
	\begin{align}
		\kappa_{\star} - \frac{1}{2}\e{\frac{m_1+\omega m_2}{\omega m_2}}(C_1^2+C_2^2+C_3^2)>0,
	\end{align}
	holds. Observe carefully that, in the special case of isotropy, it can be arranged that $\kappa>0$ for all $t\in(-\infty,\infty)$ provided $\kappa_{\star}>0$ and $m_1+m_2\omega>0$.

	\section{Perturbations and stability}
	\label{Sec:Perturbations_and_stability}
	The purpose of this section is to investigate perturbations of our exact solutions. Recall that, when deriving our solutions, one of the key assumptions we made was $\mathcal{K}_\star=0$ (see \Eqref{Eq:K_of_phi}). This restriction was made as it ensured that the resulting equation for $\phi$ was integrable. The purpose of our stability analysis here is to assess the validity of this assumption. For this reason, we focus exclusively on perturbations of the mean curvature $K$. This setting can be understood as exploring the impact of small $\mathcal{K}_\star$. While $|\mathcal{K}_\star|\gg1$ would also interesting it cannot be understood as providing information about the stability properties of our exact solutions and is therefore not discussed here.	In \Sectionref{Sec:Spatially_homogeneous_perturbations}, we begin by investigating spatially homogeneous perturbations. In this setting we find that the oscillatory solution (described in \Sectionref{Sec:Case_1}) is \emph{stable} to sufficiently small perturbations of the mean curvature. Conversely, for all cases in which a crushing singularity occurs, we find that the corresponding solutions are \emph{unstable} to perturbation, with the singularity occurring in finite proper time (and, in particular, not at $t=\pm\infty$). Spatially \emph{inhomogeneous} perturbations are then discussed in \Sectionref{Sec:Spatially_inhomogeneous_mean_curvature_perturbations}. In this setting we find that the oscillatory solution exhibits growth of its Fourier modes. Conversely, for all cases in which a Big Crunch singularity occurs, we find that the corresponding solutions are \emph{stable} to perturbation, with the singularity occurring in infinite proper time. A tabulated summary of the solutions stability properties can be found in Appendix~\ref{Sec:Summary_of_our_findings}.

	\subsection{Mean curvature perturbations with $\mathcal{K}_{\star}\ne0$}
	\label{Sec:Spatially_homogeneous_perturbations}
	
	We first consider spatially homogeneous perturbations (of our exact solutions). In this setting we are able to use various techniques from asymptotic analysis to construct a uniform approximation of the perturbed solutions. As stated above, our focus is on mean curvature perturbations only. In particular, we still require that $\nu=\omega K$ and hence $K$ is given as in \Eqref{Eq:K_of_phi}. Recall now that the exact solutions, presented above, were determined in the special case $\mathcal{K}_{\star}=0$ (see \Eqref{Eq:K_of_phi}). The goal of this subsection here is to investigate what effect a small but non-zero $\mathcal{K}_{\star}$ has on the dynamical behaviour of our exact solutions. In this case, all of our unknowns depend explicitly on $\phi$ (with the exception of $p$) and as such we only investigate what impact $\mathcal{K}_{\star}$ has on the scalar field dynamics. Let us recall now that, as a consequence of  \Eqsref{Eq:nu_particular} and \eqref{Eq:K_of_phi}, we have 
	\begin{subequations}
		\begin{align}
			\partial_{t}^{2}\phi=\frac{3}{2}\left( m_1 + m_2\omega  \right)\omega^2 + 3m_2\omega^2\phi+\frac{3}{2}\omega \mathcal{K}_\star \exp\left( \frac{2}{\omega}\phi \right).
			\label{Eq:2ndOrder_phiEq}
		\end{align}
		This equation was obtained by differentiating the evolution equation for $\phi$. A more convenient form can be obtained by defining
		\begin{align}
			\psi = \phi + \phi_\star,
			\quad
			\epsilon = \frac{3}{2}\omega \mathcal{K}_\star\exp\left( -\frac{m_1+m_2\omega}{m_2\omega} \right),
		\end{align}
		in which case we get 
		\begin{align}
			\partial_{t}^{2}\psi - 3m_{2}\omega^2\psi = \epsilon\exp\left( \frac{2}{\omega}\psi \right). 
			\label{Eq:2ndOrder_psiEq}
		\end{align}
		Given that $\phi$ is known to satisfy a \emph{first order} evolution equation, we must also have 
		\begin{align}
			\left( \partial_{t}\psi \right)^2 = 3m_{2}\omega^2\psi^2 - \frac{3m_\star}{4m_{2}}+\epsilon\e{2\psi/\omega}.
			\label{Eq:1stOrder_psiEq}
		\end{align} 
		We now investigate solutions of this equation under the assumption that $\epsilon\ll1$ (and hence $\mathcal{K}_\star\ll1$). In what follows, we consider each case separately. For our discussions here we restrict our attention to $\mathcal{K}_\star>0\implies\epsilon>0$. This is not a significant restriction, and we emphasize that similar results can be derived for $\mathcal{K}_\star<0$. Once a uniform approximation (for $\psi$) has been established we then compare our approximation to numerical solutions of \Eqref{Eq:2ndOrder_psiEq}. Note here that, in this subsection only, we do not yet make use of the evolution code discussed in \Sectionref{Sec:NumericalSetUp}. Instead, we solve \Eqref{Eq:2ndOrder_psiEq} directly using the adaptive \emph{SciPy} integrator\footnote{See \url{https://docs.scipy.org/doc/scipy/reference/generated/scipy.integrate.odeint.html}.} \emph{odeint} with an absolute error tolerance of $10^{-12}$. 
	\end{subequations}

	\subsubsection{Case 1: The oscillatory universe} 
	Let us first suppose that $(m_2<0,m_\star>0)$. We have seen already that, in this regime, solutions are oscillatory (see \Sectionref{Sec:Case_1}). It is well known that, for equations of this type, regular perturbation methods fail. Thus, it is necessary to use the \emph{method of strained coordinates} (see, for example, \cite{MultipleScales}). To this end, we write 
	\begin{subequations}
		\label{Eq:StrainedCoordinates}
		\begin{align}
			\psi(\tau) = \lambda\cos(\tau) + \epsilon\, \psi_{1}(\tau) + O(\epsilon^2), 
		\end{align}
		with 
		\begin{align}
			\tau = ( \omega_\star + \omega_1\,\epsilon + O(\epsilon^2) )t,
		\end{align}	
	\end{subequations}
	where $\psi_{1}(\tau)$ is an unknown function and $\omega_1$ is a freely specifiable constant. Inserting \Eqref{Eq:StrainedCoordinates} into \Eqref{Eq:2ndOrder_phiEq}, expanding about $\epsilon=0$, and retaining the $O(\epsilon)$-terms now gives
	\begin{subequations}
		\begin{align}
			\omega_{\star}^2(\partial_{\tau}^{2}\psi_1 + \psi_1) = 2\omega_{\star}\omega_1\lambda\cos( \tau ) + \e{\lambda\cos(\tau)/\omega}.
			\label{Eq:order1_psi}
		\end{align}
		To progress we once again make use of the Jacobi-Anger type expansion given in \Eqref{Eq:JacobiAngler_Case_1}. Note now that the homogeneous\footnote{Note here that we are using term \emph{homogeneous}, not to regards to the spatial properties, but instead to mean ``the solution without the presence of the source-terms''.} solution of \Eqref{Eq:order1_psi} is $\lambda\cos(\tau)$. Thus, if the right-hand-side of \Eqref{Eq:order1_psi} contains a $\cos(\tau)$ pieces, a secular term will be produced. With this in mind, we set 
		\begin{align}
			\omega_1=-\frac{1}{\lambda\omega_{\star}}I_{1}\left( \frac{\lambda}{\omega} \right).
		\end{align}
		Since $I_{1}(z)>0$ for all $z>0$, it follows that $\omega_1<0$. Consequently, the non-linear coupling strictly decreases the oscillation frequency relative to the linearized approximation. We note briefly that this has the consequence of increasing the observed blueshift discussed at the end of \Sectionref{Sec:Case_1}. 
		
		Using this in the evolution equation \Eqref{Eq:order1_psi} and integrating the resulting expression gives
		\begin{align}
			\label{Eq:psi_1}
			\psi_{1} = \frac{1}{\omega_{\star}^2}I_{0}\left( \frac{\lambda}{\omega} \right) + \psi_{\star}\cos(\tau) + 2\sum_{n=2}^{\infty}\frac{1}{(1-n^2)\omega_{\star}^2}I_{n}\left( \frac{\lambda}{\omega} \right)\cos(n\tau),
		\end{align}
		where $\psi_{\star}\in\mathbb{R}$ is an integration constant. 
	\end{subequations}
	\begin{figure}[t!]
		\centering
		\includegraphics[width=0.65\linewidth]{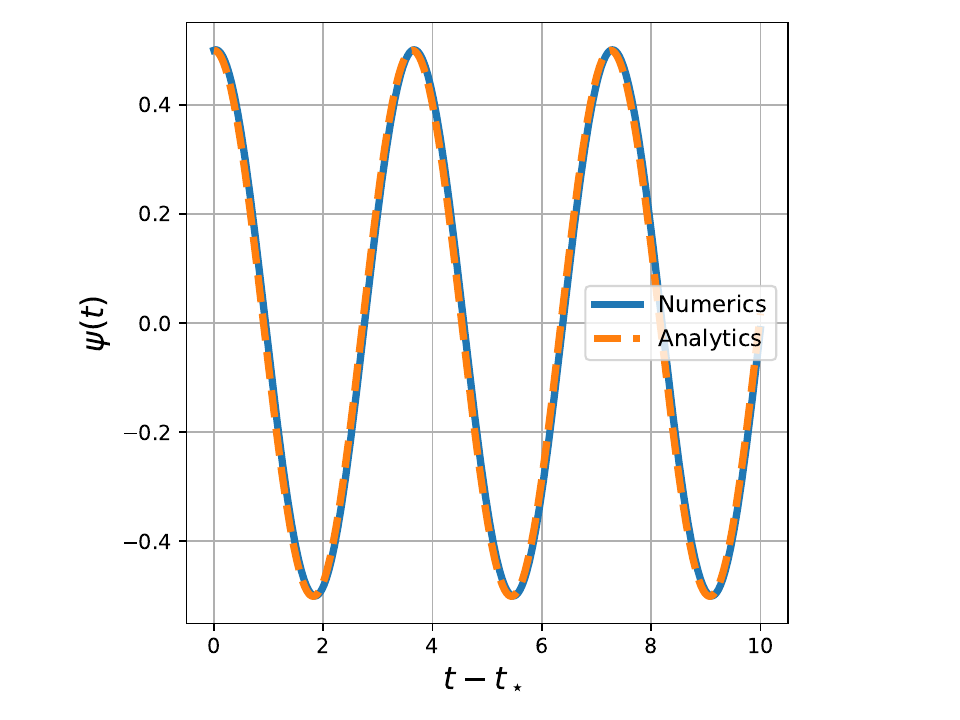}
		\caption{Numerical solution of \Eqref{Eq:2ndOrder_phiEq}, compared to the analytic approximation, in the case $m_0,m_1,m_2,\omega=0.25,1,-1,1$ and $\epsilon=10^{-3}$. In order to plot the analytical solution given in \Eqref{Eq:psi_1} we must truncate the sum. In this plot we include the first $50$ terms only. Moreover, $\psi_{\star}$ is chosen to ensure that \Eqref{Eq:1stOrder_psiEq} holds true at the initial time. }
		\label{fig:peturbcos}
	\end{figure}
	From here, we find that the results presented in \Sectionref{Sec:Case_1} are largely the same (with corrections to the period).	We therefore conclude that, at least for sufficiently small $\epsilon$, the oscillatory is stable under perturbation. A comparison of our perturbation expansion and an exact numerical solution is provided in \Figref{fig:peturbcos}.
	
	\subsubsection{Case 2: Bracketed by singularities}
	\label{Sec:Perturb_Case_2} 
\begin{figure}
		\centering
		\includegraphics[width=0.65\linewidth]{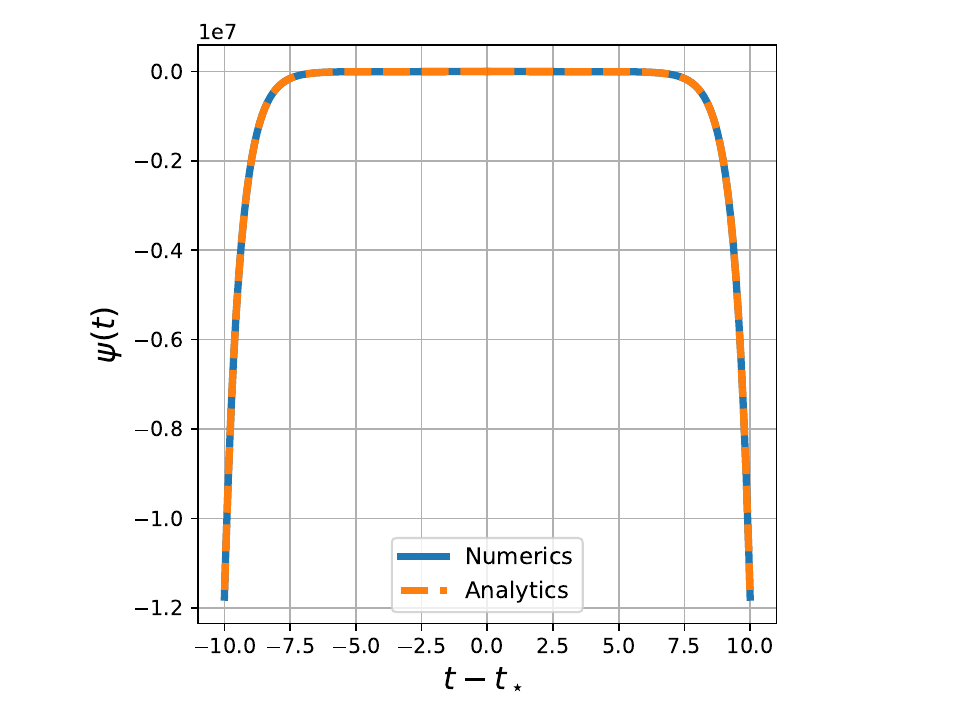}
		\caption{Numerical solution of \Eqref{Eq:2ndOrder_phiEq} (in the case $\psi<0$ as $|t|\rightarrow\infty$), compared to the analytic approximation \Eqref{Eq:psi_pertub_cosh_outer}, in the case $m_0,m_1,m_2,\omega=0.25,1,-1,1$ and $\epsilon=10^{-4}$.}
		\label{fig:peturbcoshdecay}
	\end{figure}

	We now consider the case $(m_2>0,m_\star>0)$. For this, we begin with a standard perturbation expansion to find that
	\begin{align}
		\psi = \pm\lambda\cosh(\omega_{\star}(t-t_\star)) + O(\epsilon).
		\label{Eq:psi_pertub_cosh_outer}
	\end{align}
	The validity of this expansion depends on the sign. If a minus sign is chosen then the dynamics suggested by \Eqref{Eq:2ndOrder_psiEq} is dominate the linear term and hence we expect this expansion to remain valid for all $t\in(-\infty,\infty)$. In fact, the error of this approximation decreases super exponentially as $|t|\rightarrow\infty$, with the largest error occurring in a neighbourhood of $t_\star$. In \Figref{fig:peturbcoshdecay} we provide a comparison of the expansion \Eqref{Eq:psi_pertub_cosh_outer} to a numerical solution (corresponding to the minus sign). This suggests that the Big Rip type singularities (which occur in infinite proper time) are stable to perturbations of this type.  
	
	Conversely, if the positive sign is chosen then the dynamics becomes more complicated and it is necessary to apply the \emph{method of matched asymptotics} (which splits the problem into ``inner'' and ``outer'' regions \cite{BoundaryLayer}). To understand why we observe that, in this setting, \Eqref{Eq:2ndOrder_psiEq} contains two competing terms: A linear and an exponential one. In a neighbourhood of $t=t_\star$ (the ``outer'' region) the dynamics is dominated by the linear term and hence \Eqref{Eq:psi_pertub_cosh_outer} is expected to be valid. Further away from $t_\star$, $\psi$ becomes large. When $\psi\sim-\omega\ln(\epsilon)$, the dynamics is dominated by the exponential term (this is the ``inner'' region). The change in dominant balance occurs at some time $t=t_c$, which can be determined by the equation 
	\begin{align}
		\left|3m_{2}\omega^2\psi_{(o)}(t_c)^2\right| = \left|\epsilon\exp\left( \frac{2}{\omega}\psi_{(o)}(t_c) \right)\right|,
		\label{Eq:DomiantBalance}
	\end{align}
	where $\psi_{(o)}$ is the solution of the outer problem which, in this case, is given by \Eqref{Eq:psi_pertub_cosh_outer}. Note that \Eqref{Eq:DomiantBalance} follows from \Eqref{Eq:1stOrder_psiEq}. Since the outer solution is an even function we note that $t=-t_c$ is also a solution \Eqref{Eq:DomiantBalance}. We discuss $t_c$ and $-t_c$ separately, beginning with the positive case.

	In order to investigate the dynamics of the solution for $t\ge t_c>t_\star$ we define a shifted unknown and a ``zoomed in'' time coordinate $\tau$ as
	\begin{subequations}
		\begin{align}
			\tau = \frac{t_c-t-t_\star}{\sqrt{\epsilon}},
			\quad
			\psi =-\omega\ln(\epsilon) + \Psi + O(\epsilon\ln(\epsilon)). 
			\label{Eq:Zoomin}
		\end{align}
		Plugging these into \Eqref{Eq:1stOrder_psiEq}, multiplying the resulting expression by $\epsilon$, and taking the limit $\epsilon\rightarrow0$ gives the \emph{inner problem} as 
		\begin{align}
			(\partial_{\tau}\Psi)^2 = \e{2\Psi/\omega}\implies \Psi = -\omega\ln\left( \frac{\tau}{{\omega}}+\tau_\star \right) = -\omega\ln\left( \frac{t_c - t-t_\star}{\omega\sqrt{\epsilon}} + \tau_\star \right),
		\end{align} 
		where $\tau_\star\in\mathbb{R}$ is an integration constant which is determined by requiring that the inner and outer solutions match in the limit $t\rightarrow t_c$. Direct calculation reveals
		\begin{align}
			\tau_{\star} = t_\star +  \frac{1}{\epsilon}\exp\left( - \frac{\lambda}{\omega}\cosh(\omega_{\star}(t_c-t_\star))\right).
			\label{Eq:taustar}
		\end{align}
	\end{subequations}
	A similar approach can be followed for $t\le-t_c<t_\star$. Following the procedure we find that 
	\begin{align}
		\Psi = -\omega\ln\left( \frac{t_c + (t-t_\star)}{\omega\sqrt{\epsilon}}  +\tau_\star \right),
	\end{align}
	where $\tau_\star$ is defined as in \Eqsref{Eq:taustar}. We put all of this together to find the composite solution as 
	\begin{subequations}
		\begin{align}
			\begin{split}
				\psi_\pm = \lambda\cosh( \omega_{\star}(t-t_\star) ) -\omega\ln\left( \frac{t_c \pm t}{\omega\tau_\star\sqrt{\epsilon}} + 1 \right) + O(\epsilon\ln(\epsilon)).
			\end{split}
			\label{Eq:Composite_1}
		\end{align}
		Observe carefully that, unlike our exact background solution, our composite solution exhibits a singularity at the finite times
		\begin{align}
			t_{\pm}=\pm t_{c} + \sqrt{\epsilon\omega}\tau_{\star}=\pm t_{c}+\frac{\sqrt{\omega}}{\sqrt{\epsilon}}\exp\left( - \frac{\lambda}{\omega}\cosh(\omega_{\star}(t_c-t_\star))\right).
		\end{align}
		It should be emphasised that $t=t_{\pm}$ is a physical singularity: We have already seen that if the scalar field becomes infinite, so does the Kretschmann scalar. Now, since the exponential term (in \Eqref{Eq:1stOrder_psiEq}) is exponentially suppressed for $t\in(-t_c,t_c)$, the logarithmic correction becomes relevant only once the dominant balance is reached. Consequently we construct the uniformly valid approximation
		\begin{align}
			\psi =
			\begin{cases}
				\psi_{-}, \quad &t_-<t \le -t_c,
				\\
				\psi, \quad &-t_c<t<t_c,
				\\
				\psi_+, \quad &t_c\le t< t_+. 
			\end{cases}
		\end{align}
	\end{subequations}
	\begin{figure}[t!]
		\centering
		\includegraphics[width=0.65\linewidth]{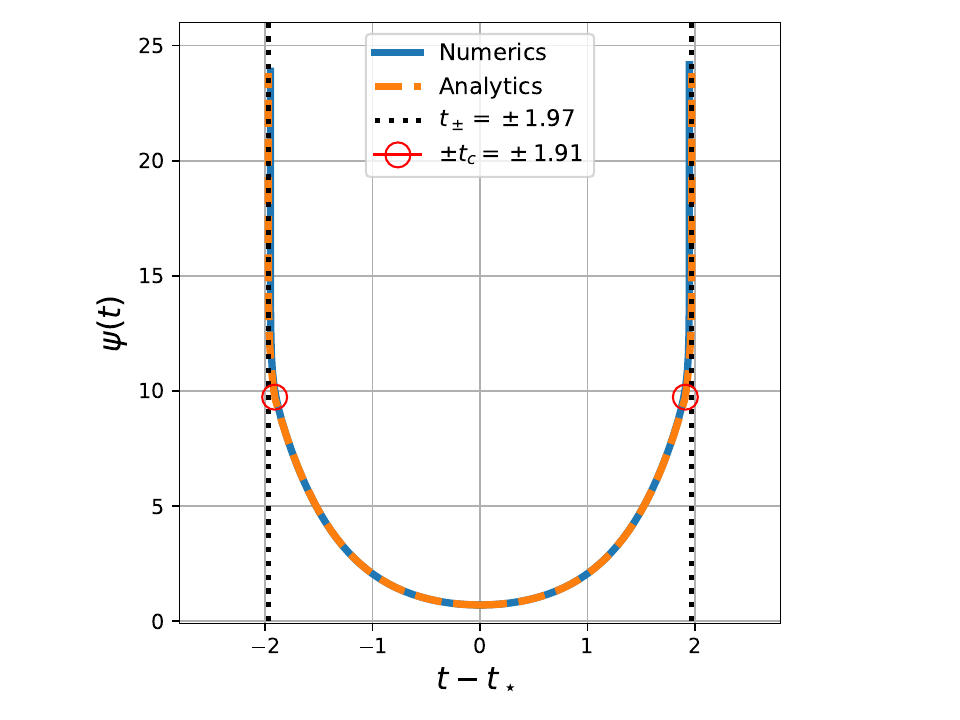}
		\caption{Numerical solution of \Eqref{Eq:2ndOrder_psiEq} compared to the composite solution in the case $m_{2},m_{1},m_{0}=1,2,1/4,\omega=1$ and $\epsilon=10^{-6}$. For these parameter values we calculated $t_c$ numerically by solving \Eqref{Eq:DomiantBalance} using the bisection method. }
		\label{fig:pic}
	\end{figure}
	In \Figref{fig:pic} we compare our analytical approximation with the numerically obtained solution in the case\footnote{Observe carefully that this particular background does not satisfy the null energy condition. Satisfying this condition is not necessary for the asymptotic approximation discussed here.} $m_{2},m_{1},m_{0}=1,2,1/4,\omega=1$ and $\epsilon=10^{-6}$. Here we find that our analytics provide a reasonably accurate description of the solutions dynamics. Note that, in this particular case we find that $t_{\pm}-t_\star\approx 1.97$.  
	
	Given all of this we conclude that this branch of solutions are \emph{not stable} to perturbations of the mean curvature as the asymptotic structure is not preserved with the singularities (formally located at $t=\pm\infty$) now becoming significant in finite proper time. 
	
	We now end this subsection by briefly discussing the behaviour of the solution near the past singularity. For this, we define $T = t_- - t$. Then, \Eqsref{Eq:K_of_phi} and \eqref{Eq:Composite_1} imply that, near $T=0$
	\begin{subequations}
		\label{Eq:PastSingulairty}
		\begin{align}
			\phi(T) \sim \phi_{\star\star} +  \phi_\star \ln(T) + \dots, 
			\quad
			K(T) \sim  -T^{-1} + \dots,
		\end{align}
		while \Eqsref{Eq:kappa} and \eqref{Eq:p_of_phi} imply that
		\begin{align}
			\kappa\sim \kappa_{\star} + \kappa_{\star\star}T^{-1} + \dots,
			\quad
			p=p_\star \ln(T) + \dots,
		\end{align}
		where $\phi_\star,\phi_{\star\star},\kappa_\star,\kappa_{\star\star},p_\star$ are constants (the values of which are not important here). This suggests that the approach to the singularity is Kasner-like. Although it should be noted that, unlike standard Kasner type singularities, there is no constraint of the squares of  ``Kasner exponents'' (which are related to the anisotropy constants $C_i$). This is consistent with results established in \cite{Ritchie:2026}, and we direct the interested reader their for a more detailed discussion of Bianchi I Kasner-like singularities in this near-minimal setting.
	\end{subequations}

	\subsubsection{Case 3: A Big Bang and a Big Rip} 
	\begin{figure}[t!]
		\centering
		\includegraphics[width=0.65\linewidth]{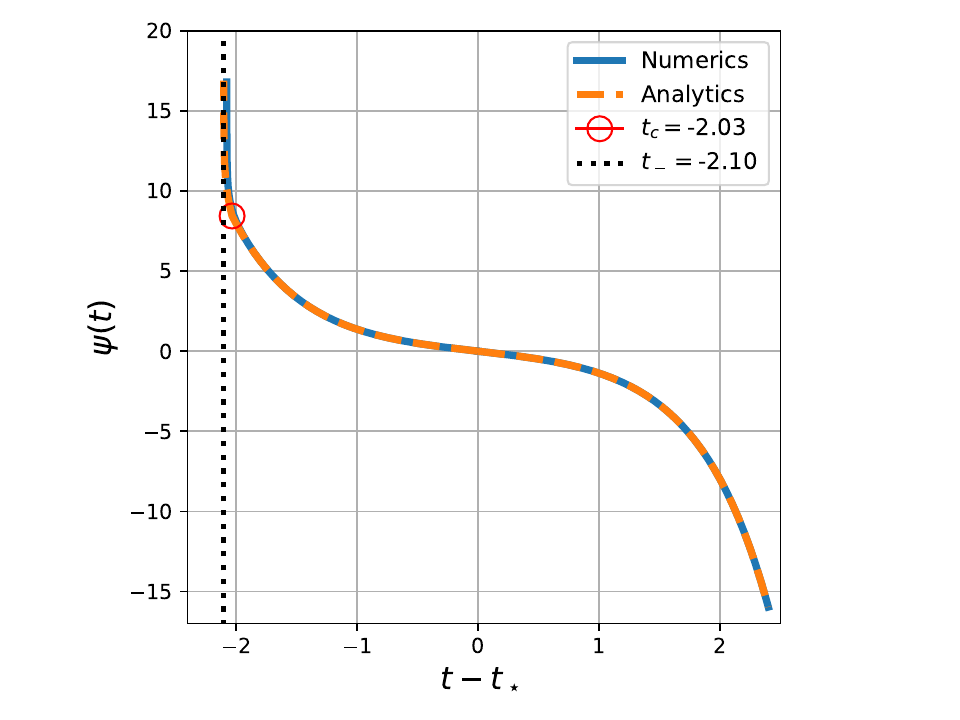}
		\caption{Numerical solution of \Eqref{Eq:2ndOrder_psiEq} compared to the composite solution in the case $m_{2},m_{1},m_{0},\omega=1,2,1,1$ and $\epsilon=10^{-4}$. For these parameter values we calculated $t_c$ numerically by solving \Eqref{Eq:DomiantBalance} using the bisection method.}
		\label{fig:peturbsinhdecay}
	\end{figure}
	We move on now to the parameter regime $(m_2>0, m_\star<0)$. As with our previous case we begin with a standard perturbation procedure
	\begin{align}
		\psi = \pm\lambda\sinh(\omega_{\star}(t-t_\star)) + O(\epsilon).
		\label{Eq:psi_pertub_sinh_outer}
	\end{align}
	Here, the sign does not play a significant role (other than to reverse the dynamics). As such, we focus only on the minus sign. In this case we find that, as $t\rightarrow\infty$ the dynamics suggested by \Eqref{Eq:2ndOrder_psiEq} is dominated by the linear term and hence we expect this expansion to remain valid for all $t\in\left[t_\star,\infty\right)$. Conversely, as $t\rightarrow-\infty$, $\psi\rightarrow\infty$ and hence the dynamics becomes dominated the exponential term (which happens when $\psi\sim-\omega\ln(\epsilon)$). The change in dominant balance occurs at a time $t_c$, defined by the dominant balance equation \Eqref{Eq:DomiantBalance} with $\psi_{(o)}$ given by \Eqref{Eq:psi_pertub_sinh_outer}. The procedure for approximating $\psi$ in the ``inner region'' $t<t_c$ is identical to what was described in above in \Sectionref{Sec:Perturb_Case_2}. Carrying out the procedure gives the uniformly valid approximation
	\begin{subequations}
		\begin{align}
			\psi = \begin{cases}
				-\lambda\sinh( \omega_{\star}(t-t_\star) ) + O(\epsilon), \quad &t \ge t_c,
				\\
				-\lambda\sinh( \omega_{\star}(t-t_\star) ) -\omega\ln\left( \frac{t_\star-t_c - t}{\omega\tau_\star\sqrt{\epsilon}} + 1 \right) + O(\epsilon\ln(\epsilon)), \quad &t_-<t < t_c,
			\end{cases}
		\end{align}
		where 
		\begin{align}
			t_{-} = t_\star-t_c -\omega\sqrt{\epsilon}\tau_\star,
			\quad
			\tau_{\star} = t_\star +  \frac{1}{\epsilon}\exp\left(  \frac{\lambda}{\omega}\sinh(\omega_{\star}(t_c-t_\star))\right).
		\end{align}
	\end{subequations}
	We once again note that we now have a physical singularity occurring at $t=t_-$ and hence conclude that Case 3 solutions are unstable, to mean curvature perturbations, in the direction of the crushing singularity, and stable in the direction of Big Rip singularities.
	
	In \Figref{fig:peturbsinhdecay} we compare our analytical approximation with the numerically obtained solution in the case $m_{2},m_{1},m_{0},\omega=1,2,1,1$ and $\epsilon=10^{-4}$. Here, we once again find that our analytics provide a reasonably accurate description of the solutions dynamics.
	
	Formal expansions of $\phi$ and $K$ near the past singularity $t=t_-$ again take the form given in \Eqsref{Eq:PastSingulairty}. Thus, we once again find that the singularity is of Kasner-type. 
	
	\subsubsection{Case 4: Singular and static} 
	\label{Sec:Perturb_Case_4} 
	For our final case, we consider the parameter regime $(m_2>0, m_\star=0)$. We once again begin with a standard perturbation procedure 
	\begin{align}
		\psi = \exp\left( \pm\omega_\star(t-t_\star) \right) + O(\epsilon).
		\label{Eq:psi_pertub_exp_outer}
	\end{align} 
	As with our previous case we restrict our attention to the minus sign, which puts the singularity in the past. We expect this approximation to remain valid in the limit $t\rightarrow\infty$, with error exponentially decreasing after $t=t_\star$. Conversely, as $t\rightarrow-\infty$ the exponential term in \Eqref{Eq:2ndOrder_psiEq} quickly dominate the dynamics. The change in dominant balance occurs at a time $t_c$, defined by \Eqref{Eq:DomiantBalance} with $\psi_{(o)}$ given by \Eqref{Eq:psi_pertub_exp_outer}. Following the procedure outlined above gives the uniform approximation 
	\begin{subequations}
		\begin{align}
			\psi =
			\begin{cases}
				\exp\left( -\omega_\star(t-t_\star) \right) + O(\epsilon), \quad &t \ge t_c,
				\\
				\exp\left( \pm\omega_\star(t-t_\star) \right) - \omega\ln\left( \frac{t_\star-t_c - t}{\omega\tau_\star\sqrt{\epsilon}} + 1 \right) + O(\epsilon\ln(\epsilon)), \quad &t_-< t< t_c,
			\end{cases}
		\end{align}
		with
		\begin{align}
			t_{-} = t_\star-t_c -\omega\sqrt{\epsilon}\tau_\star,
			\quad
			\tau_{\star} = t_\star +  \frac{1}{\epsilon}\exp\left(  \frac{1}{\omega}\exp(-\omega_{\star}(t_c-t_\star))\right).
		\end{align}
	\end{subequations}
	\begin{figure}
		\centering
		\includegraphics[width=0.65\linewidth]{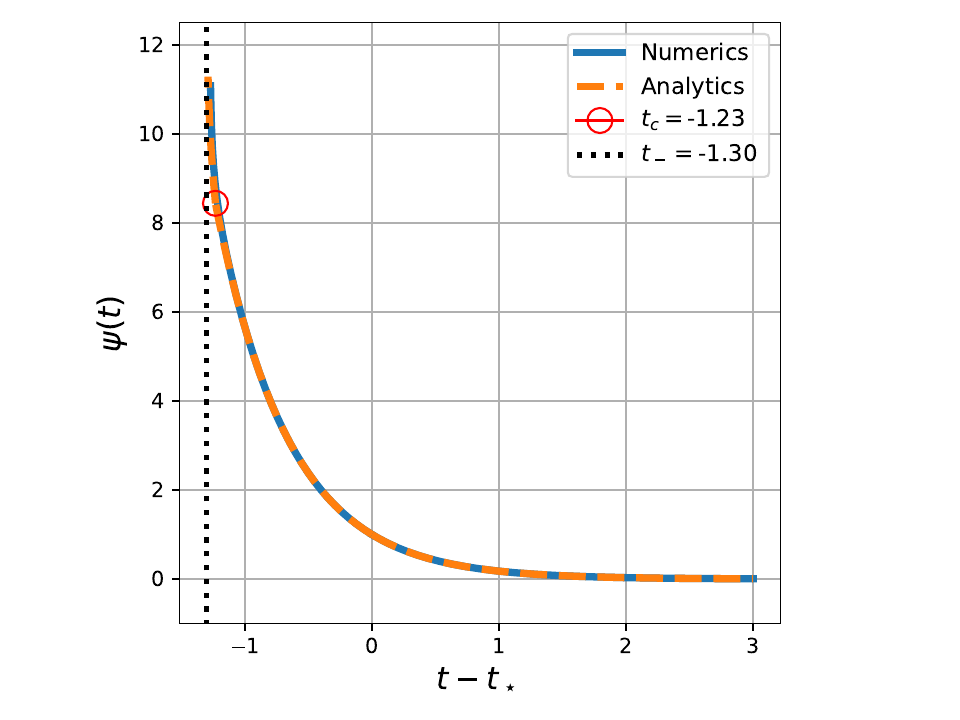}
		\caption{Numerical solution of \Eqref{Eq:2ndOrder_psiEq} compared to the composite solution in the case $m_{2},m_{1},m_{0},\omega=1,0.5,-0.1875,1$ and $\epsilon=10^{-5}$. For these parameter values we calculated $t_c$ numerically by solving \Eqref{Eq:DomiantBalance} using the bisection method.}
		\label{fig:peturbexp}
	\end{figure}
	We again find that a physical singularity now occurs at $t=t_-$ and hence conclude that Case 4 solutions are unstable, to mean curvature perturbations, in the direction of the crushing singularity, and stable in the static direction. Calculating formal expansions for $\phi$ and $K$ near the past singularity $t=t_-$ we find that they again take the form \Eqsref{Eq:PastSingulairty}, and hence this is a Kasner-type singularity. 
	
	In \Figref{fig:peturbexp} we compare our analytical approximation with the numerically obtained solution in the case $m_{2},m_{1},m_{0},\omega=1,2,1,1$ and $\epsilon=10^{-4}$. We again find that our analytical approximations provide a reasonably accurate qualitative description of the scalar fields dynamics.

	\subsection{Spatially inhomogeneous mean curvature perturbations}
	\label{Sec:Spatially_inhomogeneous_mean_curvature_perturbations}
	We now consider spatially inhomogeneous perturbations of our exact solutions. For this we make use of the code described in \Sectionref{Sec:Numerics} and hence the results presented in this subsection are purely numerical. Consequently, the conclusions presented in this subsection \emph{do not} constitute a genuine proof of stability or instability. In each of the following simulations we set $t_\star=0$ and take our initial time to be $t=0$ (or equivalently, $t=t_\star$). Now, let $(\mathring{\gamma}_{ab},\mathring{K}_{ab},\mathring{\phi},\mathring{\kappa})$ be one of our exact \emph{isotropic} solutions. Then, at the initial time $t=0$ we set 
	\begin{align}
		K(0,x,y) = \mathring{K}(0) + \delta \Phi(x,y),
		\label{Eq:MeanCurvePerturb}
	\end{align}
	for some constant $\delta\ll1$ and some freely specifiable function $\Phi(x,y)$. The constraint equations are then solved using the conformal method of Lichnerowicz and York (see Appendix~\ref{Appendix:InitialData} for more details). In doing so we obtain initial data for the fields $({\gamma}_{ab},{K}_{ab})$. The remaining fields, $\phi,\kappa,j_a$ are all chosen to match the known solution
	$(\mathring{\phi},\mathring{\kappa})$ at time $t=0$ (with $j_{a}(0)=0$). 
	
	Given such initial data we then solve the Einstein+matter equations in the increasing $t$ direction.
	Our aim here is to investigative the asymptotic behaviour $K,\phi$ and $\Omega$. Note that we have $\Omega=(\text{det}(\gamma_{ab}))^{1/6}$.
	
	To discuss this further we define the norm $\|\cdot\|_{W^{1,\infty}}$. For any scalar function $f:\mathbb{T}^2\rightarrow\mathbb{R}$ we have 
	\begin{align}
		\|f\|_{W^{1,\infty}} = \max_{x,y\in\left(0,2\pi\right]}|f| + \max_{x,y\in\left(0,2\pi\right]}\left( (\delta^{-1})^{ab}D_{a}fD_{b}f \right)^{1/2},
	\end{align}
	where $(\delta^{-1})^{ab}=\text{diag}(1,1,1)$. Our goal then is to establish bounds on the norms $\|\Omega\|_{W^{1,\infty}},$ $\|K\|_{W^{1,\infty}}$, $\|\phi\|_{W^{1,\infty}}$. In principle, this would require us to test all possible choices of the perturbation function $\Phi(x,y)$ \emph{and} all conceivable model parameters $(m_2,m_1,m_0,\omega)$. This is clearly not achievable. It should therefore be understood that the bounds we obtain here are only true for the specific cases presented, although we conjecture that they are generically true. To provide support for this claim, in each of the four cases, we consider three different choices of $(m_2,m_1,m_0,\omega)$ and four different choices of $\Phi(x,y)$. In fact, we take 
	\begin{align}
		\Phi(x,y)\in\{\sin(x)\sin(y),\sin(x)\cos(y),\cos(x)\sin(y),\cos(x)\cos(y)\}.
	\end{align}
	Finally, we note that, for each case, we shall only discuss one choice of $(m_2,m_1,m_0,\omega)$ and $\Phi(x,y)$ (in all of the simulations presented here we set $\Phi(x,y)=\sin(x)\sin(y)$). It should be understood, however, that all of the tests mentioned above have been carried out and the interested reader can find them in the supplementary material. 
	
	\subsubsection{Case 1: The oscillatory universe}
\begin{figure}[t!]
		\centering
		\includegraphics[width=1.0\linewidth]{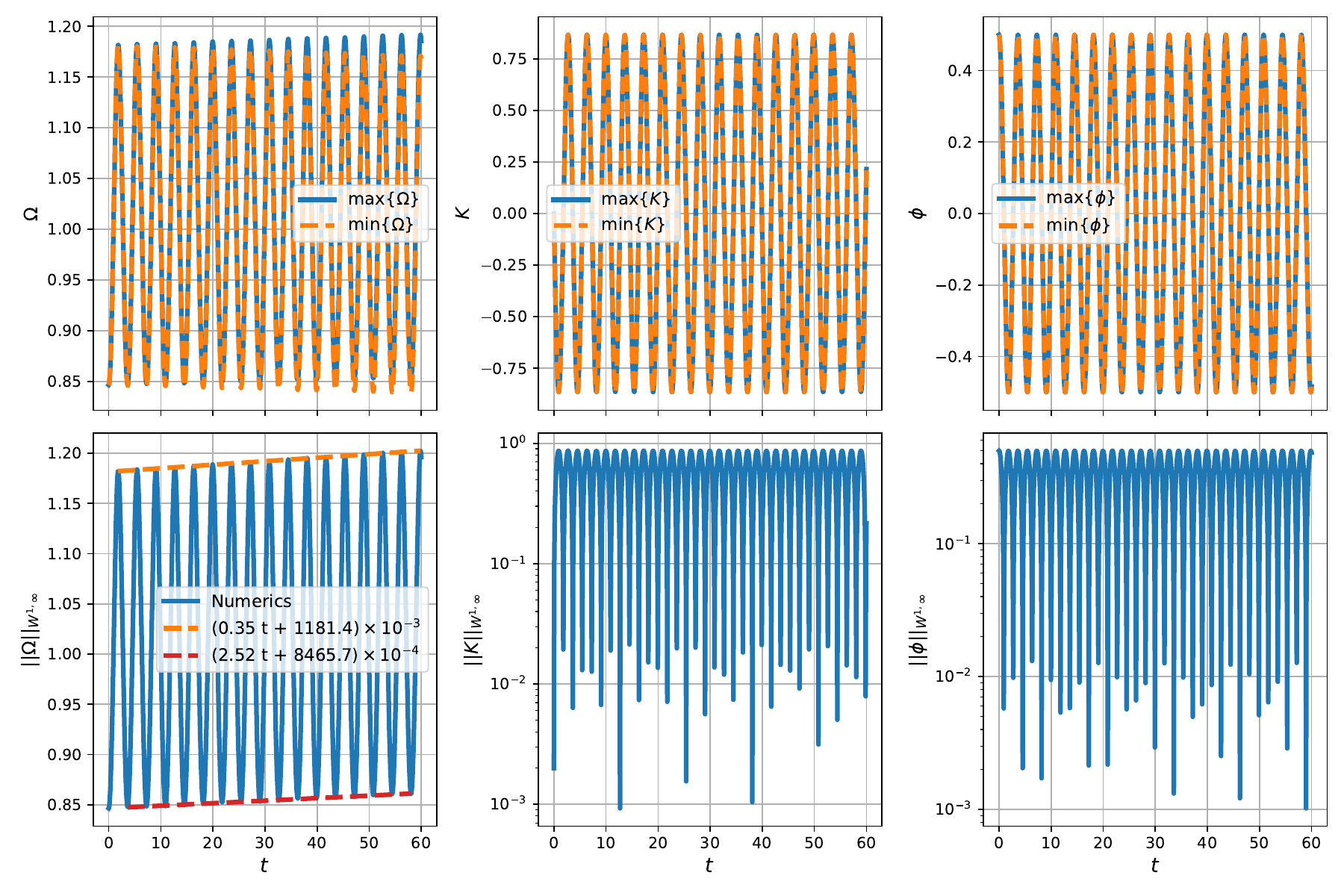}
		\caption{Numerical solutions of the Einstein scalar field system for spatially inhomogeneous mean curvature perturbations of our oscillatory solution with  $(m_0,m_1,m_2,\omega)=(0.25,1,-1,1)$, solved on the time-domain $[t_-,t_+]=[0,60]$ with $N=40$ grid points.}
		\label{fig:Case_1}
	\end{figure}

	We first consider the case defined by $(m_2<0,m_\star>0)$. In \Figref{fig:Case_1} we present the simulations corresponding to the specific choices $(m_0,m_1,m_2,\omega)=(0.25,1,-1,1)$. Our numerical results here suggest that there are constants $\mathcal{C}_1,\dots,\mathcal{C}_4>0$ such that 
	\begin{align}
		\mathcal{C}_1 t + \mathcal{C}_2 \le \|\Omega\|_{W^{1,\infty}},
		\quad
		\|K\|_{W^{1,\infty}}\le \mathcal{C}_3,
		\quad
		\|\phi\|_{W^{1,\infty}}\le \mathcal{C}_4.
	\end{align}
	i.e., the quantities $\|K\|_{W^{1,\infty}}$ and $\|\phi\|_{W^{1,\infty}}$ do not decay but instead  remain bounded for the entirety of the evolution. Conversely, the average scale factor $\Omega$ exhibits growth. 
	For our specific case we  (numerically) find that $\mathcal{C}_1\approx 2.52\times 10^{-4}$. It follows then that the spatial modes, of $\Omega$, grow without bound. Moreover, we find that $\Omega$ now exhibits long term growth (and decay) in addition to its oscillatory dynamics. This growth (and decay) is slow and it is unclear if it shall persist eternally. Whether or not these observed dynamics are due to our specific gauge choice is of course unclear. Nevertheless, our results suggest that these oscillatory solutions are not stable to spatially inhomogeneous mean curvature perturbations.

	\subsubsection{Case 2 (and 3): Stable Big Rip}
	In the previous subsection we have found that Cases 2 and 3 were unstable to spatially homogeneous mean curvature perturbations in the direction of the crushing singularity but \emph{stable} towards the Big Rip singularity. The goal now is to establish whether or not these solutions are still stable (towards  the Big Rip) in the case of spatially \emph{inhomogeneous} mean curvature perturbations. Ultimately, we find that these solutions are indeed stable in the direction of the Big Rip singularity. We note here that the criterion for demonstrating stability is different from the criterion for demonstrating instability. In the previous case, to show that the solution was unstable, it was necessary to show that at least one of the $\|\cdot\|_{W^{1,\infty}}$-norms was bounded below by a growing function. Conversely, to justify stability, it is instead necessary to show that the perturbed solution is asymptotically ```close'' to the background. For this reason, we consider the norms
	\begin{align}
		\left\|\frac{\Omega}{\mathring{\Omega}}-1\right\|_{W^{1,\infty}}\le \mathcal{C}_1,
		\quad
		\left\|\frac{K}{\mathring{K}}-1\right\|_{W^{1,\infty}}\le \mathcal{C}_{2},
		\quad
		\left\|\frac{\phi}{\mathring{\phi}}-1\right\|_{W^{1,\infty}}\le \mathcal{C}_{3},
		\label{Eq:StableEstimates}
	\end{align}
	where the ringed-quantities $(\mathring{\Omega},\mathring{K},\mathring{\phi})$ represent the perturbed background solutions.
	\begin{figure}[t!]
		\centering
		\includegraphics[width=1.0\linewidth]{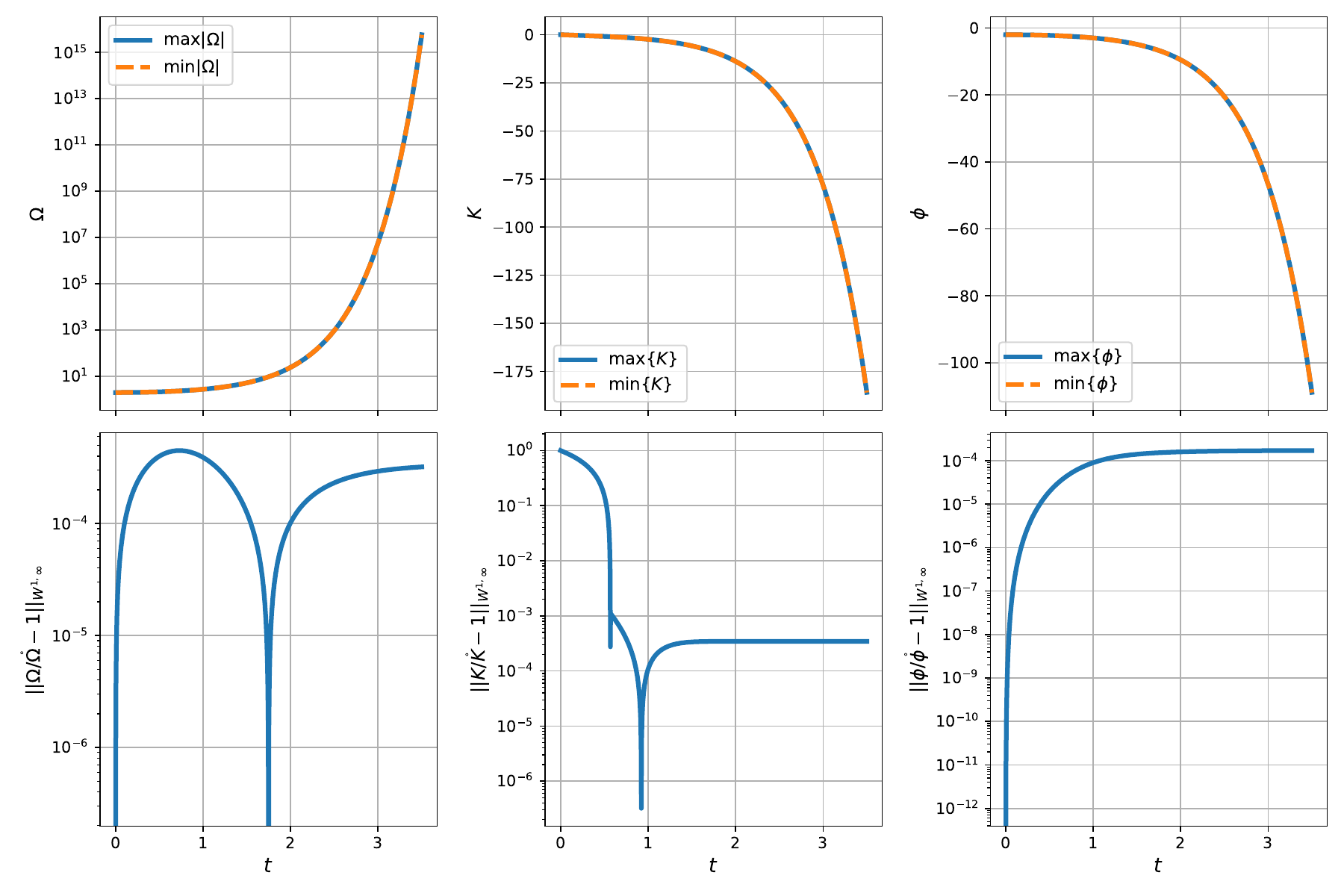}
		\caption{Numerical solutions of the Einstein scalar field system for spatially inhomogeneous mean curvature perturbations of our Case 2 solutions with  $(m_0,m_1,m_2,\omega)=(0.5,2,1,1)$, solved on the time-domain $[t_-,t_+]=[0,3.5]$ with $N=40$ grid points.}
		\label{fig:Case_2}
	\end{figure}
	In \Figref{fig:Case_2} we show the numerical results corresponding to the parameter values $(m_0,m_1,m_2,\omega)=(0.5,2,1,1)$. i.e., we consider perturbations of the ``Case 2'' solution (see \Sectionref{Sec:Case_2}). Here, we find that there are constants $\mathcal{C}_1,\mathcal{C}_2,\mathcal{C}_3>0$ that satisfy the estimates \Eqref{Eq:StableEstimates}. In fact, we find that $\mathcal{C}_1,\mathcal{C}_3=O(\delta)$. Conversely, we find $\mathcal{C}_2=O(1)$. However, it is worth noting that this is only true at the initial time. For $t>1$ we find that
	\begin{align}
		\left\|\frac{K}{\mathring{K}}-1 \right\|_{W^{1,\infty}}=O(\delta),
		\quad
		t> 1.
	\end{align}
	Ultimately, we were unable to continue these evolutions past $t=3.5$. This is likely due to that fact that, at $t=3.5$ we have $\Omega= O(10^{15})$. In regards to whether or not this is a physical singularity (and in particular not simply a consequence of our gauge) we note that, since the solutions remain ``close'' to the background, the Kretschmann scalar can be approximated by \Eqref{Eq:Kr} (at least up to $O(\delta)$) and hence the fact that $\phi$ is becoming infinite (as evidenced in \Figref{fig:Case_2}) implies that this is indeed a physical singularity. Given these results we conclude that the Case 2 solutions are stable to spatially inhomogeneous mean curvature perturbations in the direction of a Big Rip type singularity\footnote{Although these numerical results do not prove stability it is worth noting that their apparent stability is consistent with known results about the stability of Big Rip type solutions. See, for example,  \cite{Barrow:2009,Frampton:2012}, and the references therein.}. We find that this statement is also true for spatially inhomogeneous perturbations of the Case 3 solutions, discussed in \Sectionref{Sec:Case_3}. For the sake of brevity, we do not present Case 3 here. Nevertheless, the interested reader can find these simulations in the supplementary material.
	
	\subsubsection{Case 4: Singular and static}
	Finally, we consider spatially inhomogeneous perturbations of the ``Case 4'' solutions, described in \Sectionref{Sec:Case_4}. In \Sectionref{Sec:Perturb_Case_4} we found that these solutions were stable to spatially homogeneous mean curvature perturbations in the static direction and unstable in the singular direction. Our goal now is to establish whether or not this remains true in the case of spatially \emph{inhomogeneous} perturbations. 
	\begin{figure}[t!]
		\centering
		\includegraphics[width=1.0\linewidth]{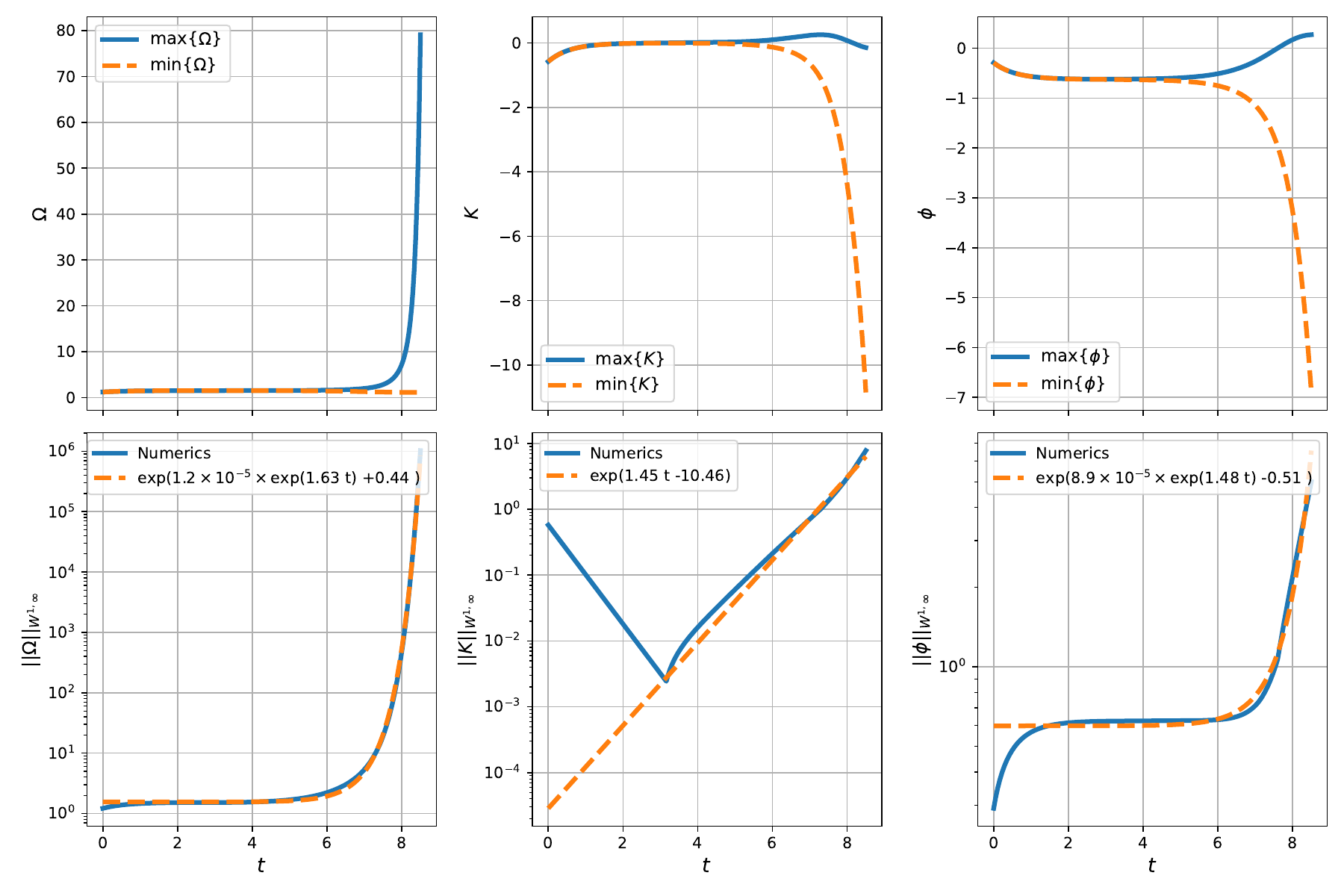}
		\caption{Numerical solutions of the Einstein scalar field system for spatially inhomogeneous mean curvature perturbations of our Case 4 solution with  $(m_0,m_1,m_2,\omega)=(-0.234375,0.25,1,1)$, solved on the time-domain $[t_-,t_+]=[0,8.5]$ with $N=40$ grid points.}
		\label{fig:Case_4}
	\end{figure}
	In \Figref{fig:Case_2} we show the numerical results corresponding to the parameter values $(m_0,m_1,m_2,\omega)=(-0.234375,0.25,1,1)$. Here, we find that there are constants $\mathcal{C}_1,\dots,\mathcal{C}_8>0$ such that 
	\begin{align}
		\|\Omega\|_{W^{1,\infty}}\ge \mathcal{C}_1\exp\left(\mathcal{C}_2\e{\mathcal{C}_3 t}\right),
		\quad
		\|K\|_{W^{1,\infty}}\ge \mathcal{C}_{4}\e{ \mathcal{C}_5 t },
		\quad
		\|\phi\|_{W^{1,\infty}}\ge \mathcal{C}_6\exp\left( \mathcal{C}_{7}\e{\mathcal{C}_8 t}\right).
	\end{align}
	For our specific case we have $\mathcal{C}_3\approx1.63,\mathcal{C}_5\approx1.3,$ and $\mathcal{C}_8\approx1.47$. We find that these norms become infinite as $t\rightarrow\infty$. In \Figref{fig:Case_2} we see that, at some spatial points, the solution appears to behave like a static universe. However, at other points the $\Omega$ is growing super-exponentially. We therefore conclude that Case 4 solutions are unstable to spatially inhomogeneous mean curvature perturbations in the static direction.

	\section{Conclusion}
	\label{Sec:Conclusion}
	
	The goal of this work has been to construct exact solutions of the Einstein equations coupled to a ``near-minimal'' scalar field. This non-variational model is defined by a wave equation with a self-interaction $f(\phi)$ that is not equal to the derivative of the gravitational potential $V(\phi)$. i.e., $f(\phi) \neq V'(\phi)$. In order to construct these solutions we restricted our attention to spatially flat, homogeneous Bianchi I geometries, and set $f(\phi) = 3\omega V(\phi)$. Moreover, we chose $V(\phi)$ to be a quadratic function. These choices were not motivated by physical considerations but instead by their mathematical tractability. Physically motivated choices of $f(\phi)$ and $V(\phi)$ remains a direction for future work. In any case, given these particular choices, we successfully derived four distinct explicit solutions of the Einstein+matter system.

	The first of these, Case 1, represents an oscillatory (cyclic) cosmology notably characterized by the absence of singularities. Conversely, all of our other solutions did possess one or more singularities. The Case 2 solution describes a universe bounded by two singularities, the nature of which depends on the asymptotic behaviour of the scalar field. On the one hand, if $\phi \rightarrow +\infty$ as $|t| \rightarrow \infty$, both the volume element tends to zero and hence singularities are of crushing type. On the other hand, if $\phi \rightarrow -\infty$, the volume element diverges and hence both singularities represent an expansionary ``Big Rip'' singularity. The Case 3 solution is structurally similar Case 2 but possesses asymmetric singularities at $t = \pm\infty$, where one of the singularities is crushing type (i.e., Big Bang or Big Crunch) and the other expansionary (Big Rip). Finally, Case 4 describes a universe that is singular in one temporal direction and static in the other. Notably, all of the singularities, present in Cases 2,3, and, 4, occur in infinite proper time and are therefore formally unreachable by an observer.
		
	In order to assess the physical viability of these models we then investigated the stability properties of our exact solutions under both spatially homogenous and spatially inhomogeneous perturbations of the mean curvature. A tabulated summary of our findings can be found in Appendix~\ref{Sec:Summary_of_our_findings}.
	
	In the spatially homogenous setting we used a mixture of numerical and analytical techniques to establish our results. Here, we found that solutions featuring crushing singularities are inherently unstable to perturbations, with the singularity shifting to finite proper time. Conversely, we found the solutions of a Big Rip singularity \emph{were} stable to perturbations of this type. Similarly, we found that the oscillatory (Case 1) solution was stable to mean curvature perturbations. In regards to the Case 4 solution, we found that it was stable to mean curvature perturbations in the static direction.
	
	In the spatially inhomogeneous setting, we use purely numerical methods to determine stability. In this setting, we found that the Case and Case 4 solutions were unstable to spatially inhomogeneous mean curvature perturbations. Conversely, solutions in possession of a Big Rip singularity were found to be stable to spatially inhomogeneous perturbations of the mean curvature. While promising, our stability results (particularly in the spatially inhomogeneous setting) have been established using numerically only and as such do not constitute a genuine proof of stability. In future works, it would be interesting to formally prove this claim; demonstrating that these Big Rip type singularities are indeed stable. Additionally, it would be interest to consider more physically motivated choices for $f(\phi)$ and $V(\phi)$. It is also unclear how the particular choices of these functions impact results such as the BKL conjecture and Strong Cosmic Censorship (see, for example, \cite{BKL:1970,Ringstrom2010,Fournodavlos:2023} for discussions on the BKL conjecture and \cite{Luna:2019,VanDeMoortel:2025,Fajman:2025} for discussions on Strong Cosmic Censorship). Investigating how the $f(\phi)$ {and} $V(\phi)$ impact the conjectures is therefore an intriguing possible direction for future research.

	\newpage
	\bibliographystyle{unsrt}
	\bibliography{bibfile}
	
	\newpage
	\begin{appendices}
		\section{Numerical infrastructure}
		\label{Appendix:Numerics}
		The purpose of this appendix is to expand on the discussion presented in \Sectionref{Sec:NumericalSetUp}. Our code must do three key things. First, it must construct initial data. For this we use the conformal method of Lichnerowicz and York; the details of which are discussed in \Sectionref{Appendix:InitialData}. Second, our code must evolve this initial data, while suppressing constraint violations. Details, and numerical tests, of our constraint dampening scheme is discussed in \Sectionref{Appendix:ConstraintDampening}. Third, our code must produce results, defined on $\tilde{M}$ (and hence $M$), that converge appropriately. Convergence tests are discussed in \Sectionref{Appendix:ConvergenceTesting}.

		\subsection{Constructing initial data}
		\label{Appendix:InitialData}
		We first discuss how we numerically solve the constraint equations \Eqsref{Eq:Ham} and \eqref{Eq:Mom}. For this we use the conformal method of Lichnerowicz and York. The primary goal of this subsection is to provide a brief summary of this method and to describe our numerical procedure for solving the results equations. The discussion presented here should be understood as a brief summary only. For a more detailed discussion of the conformal method we refer the interested reader to \cite{Alcubierre:Book} and the references therein.
		
		Consider now two quadruples $(\Sigma,\gamma_{a b},{\chi^a}_b,K)$ and $({\Sigma},\bar{\gamma}_{a b},{\bar{\chi}^a}{}_b,\bar{K})$; where $\Sigma$ is an arbitrary  $3$-dimensional differentiable manifold, with Riemannian metrics $\gamma_{a b}$ and $\bar{\gamma}_{a b}$, and $K,\bar{K}$ are smooth scalar functions, and ${\chi^a}_{b},{\bar{\chi}^a}{}_{b}$ are trace-free tensor fields. Here we set 
		\begin{align}
			{K^a}_{b}={\chi^a}_{b} + \frac{1}{3}K{\gamma^{a}}_{b},
			\quad
			{\bar{K}^a}{}_{b}={\bar{\chi}^a}{}_{b} + \frac{1}{3}\bar{K}{\bar{\gamma}^{a}}{}_{b}.
		\end{align}
		
		At this stage none of these quantities are required to satisfy any equation (such as the constraint equations). The Levi-Civita covariant derivatives associated with $\gamma_{a b}$ is $D_a$. Suppose now that there exists a smooth function $\psi:\Sigma\rightarrow\mathbb{R}$ so that the metrics $\gamma_{a b}$ and $\bar{\gamma}_{ab}$ are proportional. The particular relationship between the two metrics is 
		\begin{align}
			\gamma_{a b}=\psi^{4}\bar{\gamma}_{ab}.
		\end{align}
		The covariant derivative associated with $\bar{\gamma}_{ab}$ is ${\bar{D}}_a$. The tensor field 
		\begin{align}
			C^{a}_{bc}=2\left(\delta^{a} {}_{b}\bar{D}_{c}\ln\psi+\delta^{a} {}_{c}\bar{D}_{b}\ln\psi-\bar{\gamma}_{bc}\bar{\gamma}^{ad}\bar{D}_{d}\ln\psi\right),
		\end{align}
		is the smooth tensor that relates the covariant derivatives of the two metrics $D_{a}-\bar{D}_{a}$. In fact the field $C^{a}_{bc}$ can also be used to relate the Ricci tensors associated with $\gamma_{a b}$ and $\bar{ \gamma }_{ab}$ --which are labelled as $R_{ab}$ and $\bar{R}_{ab}$, respectively. 
		
		In regards to the trace and trace-free components of the extrinsic curvature, we write 
		\begin{align}
			K=\bar{ K },
			\quad
			\chi^{ab} = \psi^{-10}\left( \bar{ \chi }^{ab} + (\textbf{L}{V})^{ab} \right),
		\end{align}
		where $V^a$ is some unknown vector and where the differential operator $\textbf{L}$ is defined as
		\begin{align}
			(\textbf{L}{V})^{ab}=2\bar{D}^{ \left( a \right.  } { V }^{ \left. b \right) } - \frac{2}{3}\bar{ \gamma }^{ab}\bar{D}_c{ V }^c,
			\label{Eq:P1_Operator_L}
		\end{align}
		In addition, it is useful to define the vector Laplacian $\bar{\Delta}_{\bar{\textbf{L}}}{ V }^a$ as 
		\begin{align}
			\bar{\Delta}_{\bar{\textbf{L}}}{ V }^i = \bar{D}_{j}\left( (\textbf{L}{V})^{ij} \right)=\bar{ \gamma }^{jk} \bar{D}_{j}\bar{D}_{k}V^{i} + \frac{1}{3}\bar{D}^i \bar{D}_j { V }^j + \bar{ R }^{i}{}_j { V }^j.
		\end{align}
		
		Given all of this one readily shows that the Hamiltonian and momentum constraints \Eqsref{Eq:Ham} and \eqref{Eq:Mom} can be written as
		\begin{subequations}
			\begin{align}
				8\Delta_{\bar{\gamma}}\psi-\psi
				\bar{R}-\frac{2}{3}\psi^{5}\bar{K}^{2}+\psi^{-7}{\chi}^2&=-\psi^5 \rho,
				\label{Eq:P1_ConformalHamiltonian}
				\\
				\bar{\Delta}_{\bar{\textbf{L}}}\bar{ V }^a + {\bar{D}}_{b}{\bar{ \chi } }^{ab} -\frac{2}{ 3}\psi^{-6}\bar{D}^{ a }{ \bar{K} } &= \psi^{10} {j}^{a}.
				\label{Eq:P1_ConformalMomentum}
			\end{align}
			with
			\begin{align}
				{\chi}^2 = \bar{\gamma}_{ab}\bar{\gamma}_{cd}(\bar{ \chi }^{ac}+(\textbf{L}{V})^{ac})(\bar{ \chi }^{bd}+(\textbf{L}{V})^{bd}).
			\end{align}
		\end{subequations}
		\newcommand{\ConformalConstraints}{\Eqsref{Eq:P1_ConformalHamiltonian}--\eqref{Eq:P1_ConformalMomentum}}
		
		\ConformalConstraints{} suggest the grouping of the various fields, introduced above, as follows:
		\begin{description}
			\item[Free data:] The fields ${\rho}$, ${j}_{a}$, $\bar{ \chi }^{ab}$,  $\bar{ \gamma }_{ab}$ and $\bar{ K }$ are considered as {freely specifiable} everywhere
			on $\Sigma$. All of $\bar{D}$, $\bar{R}_{ab}$, $\bar{\Delta}_{\bar{\textbf{L}}}$ and $\bar{ R }$ (together with all of the index versions of these) in \ConformalConstraints{} are fully determined by these on $\Sigma$.
			\item[Unknowns:] The quantities $\psi$, $\bar{V}^i$ are considered as the unknowns of \ConformalConstraints{} once the free data have been specified.  
		\end{description}
		
		We now discuss how we numerically solve \ConformalConstraints. Consider the initial data set $(\Sigma,\bar{\gamma}_{ab},{\bar{\chi}^a}{}_{b},\bar{K},\bar{\phi},\bar{\kappa},\bar{j})$ which is not necessarily a solution of the constraints \Eqsref{Eq:Ham} and \eqref{Eq:Mom}. Then we, define 
		\begin{align}
			\mathring{H} = \bar{\kappa} + V(\bar{\phi}) - \frac{1}{2}( \bar{R} + \bar{K}^{2} - \bar{K}_{ab}\bar{K}^{ab} ),
			\quad
			\mathring{M}_{a} = \bar{j}_{a} + \bar{D}^{b}\bar{K}_{ab} - \bar{D}_{a}\bar{K}.
		\end{align}
		If the set $(\Sigma,\bar{\gamma}_{ab},{\bar{\chi}^a}{}_{b},\bar{K},\bar{\phi},\bar{\kappa},\bar{j})$ \emph{is} a solution of the constraints then $\mathring{H}=0,\mathring{M}_a=0$. In general, this is not the case. However, we nevertheless assume that $(\mathring{H},\mathring{M}_a)$ are somehow ``small'' and hence are ``almost a solution''. Let $\delta\ll1$ be a small constant and suppose that $H,M_1,M_2=O(\delta)$. Then we write
		\begin{align}
			\label{Eq:ConformalSols}
			\psi = 1 + \delta\hat{\psi} + O(\delta^2),
			\quad
			V^a = \delta\hat{V}^a + O(\delta^2),
		\end{align}
		for some function $\hat{\psi}$ and some vector $\hat{V}^a$. Inputting these expressions into \ConformalConstraints, expanding about $\delta=0$ now gives 
		\begin{subequations}
			\begin{align}
				H = \mathring{H} + \delta\left( 8\Delta_{\bar{\gamma}}\hat{\psi} - \left(\bar{R} + \frac{10}{3}\bar{K}^2 + 7{\bar{\chi}^a}{}_b{\bar{\chi}^b}{}_a\right) \hat{\psi} + 2\mathring{\chi}^{ab}(\textbf{L}\hat{V})_{ab} \right) + O(\delta^2),
				\\
				M_a = \mathring{M}_a + \delta\left( \bar{\Delta}_{\mathbf{L}}\hat{V}_a + 4\hat{\psi}\bar{D}_a \bar{K} \right) + O(\delta^2).
			\end{align}
			\label{Eq:LinearisedConformal}
		\end{subequations}
		The \emph{correction fields} $(\delta\hat{\psi},\delta\hat{V}_a)$ are then determined as a solution of the \emph{linear} problem
		\begin{subequations}
			\label{Eq:Linearised}
			\begin{align}
				8\Delta_{\bar{\gamma}}(\delta\hat{\psi}) - \left(\bar{R} + \frac{10}{3}\bar{K}^2 + 7{\bar{\chi}^a}{}_b{\bar{\chi}^b}{}_a\right) \delta\hat{\psi} + 2\bar{\chi}^{ab}(\textbf{L}(\delta\hat{V}))_{ab}  = -\mathring{H},
				\\
				\bar{\Delta}_{\mathbf{L}}(\delta\hat{V}_a) + 4(\delta\hat{\psi})\bar{D}_a \bar{K}=-\mathring{M}_a.
			\end{align}
		\end{subequations}
		Note that these equations can be obtained by neglecting the $O(\delta^2)$-terms in \Eqsref{Eq:LinearisedConformal}. For sufficiently small $\delta$ we expect \Eqsref{Eq:ConformalSols} (with $(\hat{\psi},\hat{V}_a)$ obtained as solutions of \Eqsref{Eq:Linearised}) to provide a reasonably accurate approximation of solutions to \ConformalConstraints. 
		
		To numerically solve \Eqsref{Eq:Linearised} in our Python code we proceed as follows: First, we make particular choices of the free fields $(\bar{\gamma}_{ab},\bar{\chi}_{ab},j_a,\rho,\bar{K})$. Our choices are discussed in \Sectionref{Sec:Spatially_inhomogeneous_mean_curvature_perturbations}. We then make use of the SciPy function\footnote{See \url{https://docs.scipy.org/doc/scipy/reference/generated/scipy.sparse.linalg.gmres.html}} \emph{gmres}. This requires us to write \Eqsref{Eq:Linearised} as a linear problem of the form $Ax=b$ where $x\in\mathbb{R}^{4 N^2}$ is a vector containing all values of the unknowns $(\delta\hat{\psi},\delta\hat{V}_a)$ at points on our numerical grid, $b\in\mathbb{R}^{4 N^2}$ is the vector containing the \emph{remainders} $(\mathring{H},\mathring{M}_a)$, and where $A$ is a linear operator that, given an $x$, calculates the left-hand-side of \Eqsref{Eq:Linearised}. The \emph{gmres} works optimally when $A$ is somehow sparse. For this reason, derivatives on the left-hand-side of \Eqsref{Eq:Linearised} are calculated using a \emph{central finite differencing} scheme while the remainders $(\mathring{H},\mathring{M}_a)$ are calculated \emph{spectrally}. This hybrid approach allows us to use finite differencing to obtain the numerical solution with spectral accuracy. This procedure can then be repeated iteratively until the solutions are sufficiently accurate. For the simulations presented here we find that $3$ iterations is often sufficient to obtain an accuracy smaller than $5\times 10^{-7}$. A table summarising this information can be found in the supplementary material.

		\subsection{Constraint dampening}
		\label{Appendix:ConstraintDampening}
		\begin{figure}[t!]
		\centering
		\includegraphics[width=1.\linewidth]{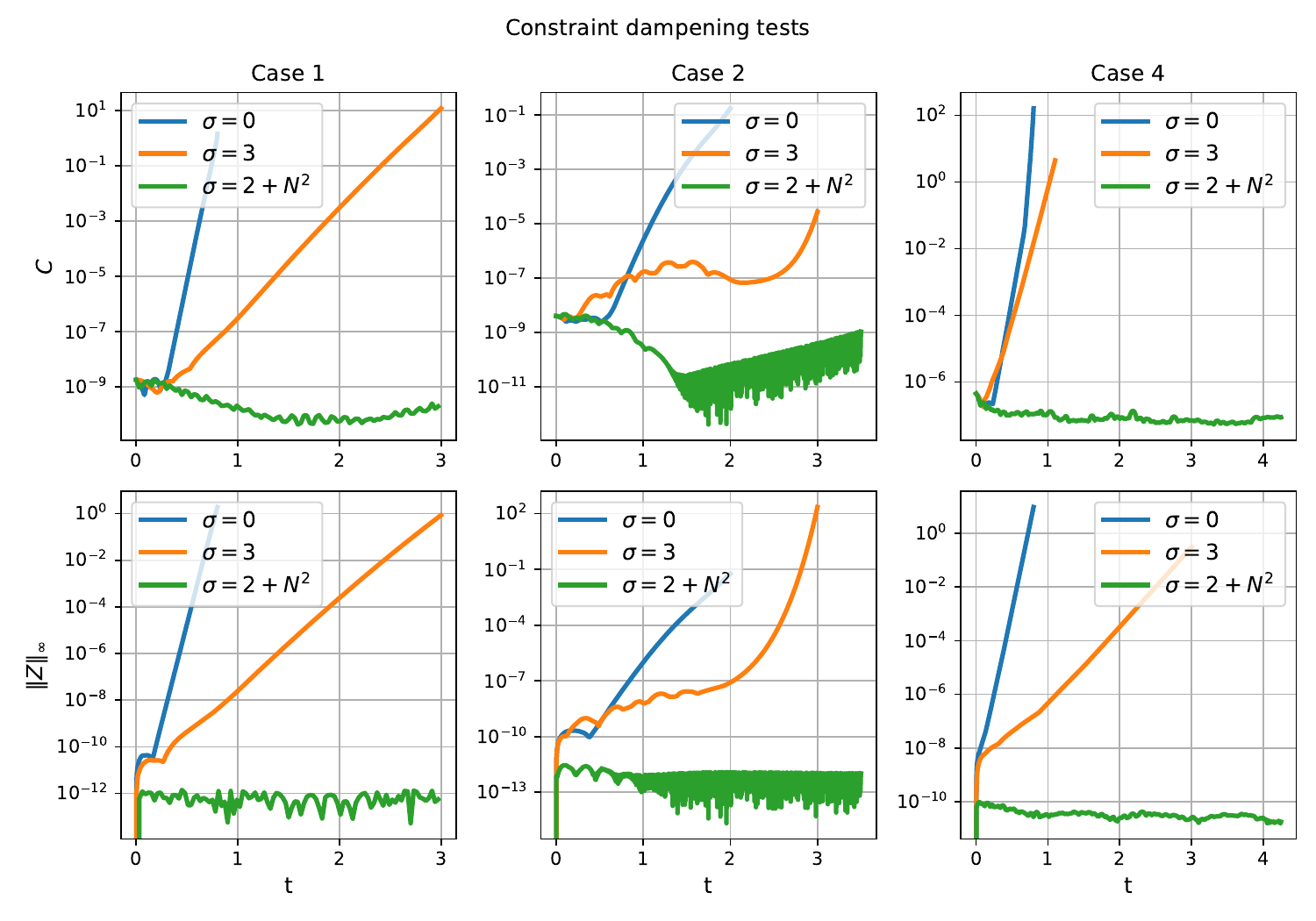}
		\caption{Constraint violations, as a function of time, for different choices of the dampening parameter $\sigma$. The top row shows the combined violations of the Hamiltonian and momentum constraints for the various simulations. Similarly, the bottom row shows the $L^\infty$ norm of the vector Z4 vector $Z^\alpha$. In regards to our model parameters, the first column shows simulations corresponding to the choices $(m_0,m_1,m_2,\omega)=(0.25,1,-1,1)$. Similarly, the second and third columns shows the results corresponding to the parameters $(m_0,m_1,m_2,\omega)=(0.5,2,1,1)$ and $(m_0,m_1,m_2,\omega)=(-0.234375,0.25,1,1)$, respectively. }
		\label{fig:CD}
		\end{figure}
		In the present work we make use of a constraint dampening scheme consistent with works such as \cite{Gundlach:2005,weyhausen2012constraint}. In \Eqsref{Eqs:FullSystem} the constant $\sigma$ represents the constraint dampening terms. When $\sigma=0$, numerical truncation errors source constraint violations, which can grow and generate numerical instability. This behaviour can be controlled by considering non-zero values of $\sigma$. We determined an appropriate value of $\sigma$ empirically. To measure the constraint violations, and hence establish the effectiveness of our constraint dampening scheme, we calculate the quantities
		\begin{subequations}
		\begin{align}
			C = \max_{x,y\in\left(0,2\pi\right]}\left( H^2 + (\delta^{-1})^{ab}M_{a}M_b \right)^{1/2},
		\end{align}
		and
		\begin{align}
			\|Z\|_{\infty} = \max_{x,y\in\left(0,2\pi\right]}\left( \Theta^2 + (\delta^{-1})^{ab}X_{a}X_b \right)^{1/2}.
		\end{align} 
		\end{subequations}
		In \Figref{fig:CD} we show the constraint violations for each of the three simulations that were presented in \Sectionref{Sec:Spatially_inhomogeneous_mean_curvature_perturbations}. We found that our constraint dampening scheme works most effectively when $\sigma$ scaled linearly with the grid resolution. For this reason, for all of our simulations, we set\footnote{It is worth noting that \cite{Gundlach:2005,weyhausen2012constraint} find that $\sigma>0$ ``should'' be sufficient for dampening constraint violations. However, as can be seen in \Figref{fig:CD}, we do not find this to be sufficient.} $\sigma=2+N^2$.

		\subsection{Error and convergence testing}
		\label{Appendix:ConvergenceTesting}		
		For given initial data (constructed as described in Appendix~\ref{Appendix:InitialData}) the Einstein+matter equations \Eqsref{Eqs:FullSystem} are solved as an initial value problem in the increasing $t$ direction starting at $t=0$. As outlined in \Sectionref{Sec:NumericalSetUp}, we employ a uniform grid with $N\times N$ points. For our time-stepping method, we use the adaptive \emph{SciPy} integrator\footnote{See \url{https://docs.scipy.org/doc/scipy/reference/generated/scipy.integrate.odeint.html}.} \emph{odeint}. We denote its absolute error as $\tilde{\mathcal{E}}$ and its magnitude as $\tilde{\varepsilon}=-\log(\tilde{\mathcal{E}})$. The parameter $\tilde{\varepsilon}$ controls the local step size of the time evolution and hence one expects that $\tilde{\varepsilon}$ and $N$ can be used to control the error that is numerically generated in our time- and space-discretisations. Let $f$ be one of the numerically calculated unknowns, corresponding to some choice of the parameters $\tilde{\varepsilon}$ and $N$. Then the absolute error $E[f](t,x,y,\tilde{\varepsilon},N)$ of $f$ can be calculated at a point $(t,x,y)$. In principle, this is done by comparing the numerically calculated unknown to the exact solution. In practice, however, the exact solution is not known, in which case we follow the common practice of determining $E[f](t,x,y,\tilde{\varepsilon},N)$ by comparing the numerical solution to \emph{another} numerical solution obtained with some sufficiently high resolution (instead of the exact solution). If $\mathring{f}$ is one such reference solution then we define the relative error, for each $f$, as 
		\begin{align}
			E_{R}[f](t) = \frac{\max_{x,y}|f-\mathring{f}|}{\max\{ 1, \|\mathring{f}\|_{\infty} \}},
			\quad
			\|\mathring{f}\|_{\infty} = \max_{x,y}|\mathring{f}|.
		\end{align}
		The reference solution $\mathring{f}$ is determined on a numerical grid with $N_2\times N_2$ points, while $f$ is determined on a grid with $N_1\times N_1$ points. In order to compare the two solutions, we use the Fourier transform to calculate $f$ on the $N_2\times N_2$ grid. 
		
		On the one hand, if $\tilde{\varepsilon}$ is sufficiently large we expect that the error is dominated by the spatial discretisation. In this setting, the error should be roughly independent of $\tilde{\varepsilon}$, but should decrease monotonically with $N$. We always set $\tilde{\varepsilon}=12$, unless stated otherwise. On the other hand, if $N$ is large enough to resolve all spatial features of the solution, then the numerical error is dominated by the time discretisation. In such a setting, the numerical error should not become smaller when we increase $N$ (in fact, oversampling may be a significant error source). The error should decrease monotonically with $\tilde{\varepsilon}$. 
		
		In order to demonstrate this behaviour it is useful now to define the \emph{total error}. For any fixed $N$ and $\tilde{\varepsilon}$ calculate
		\begin{align}
			\mathcal{E}(t)=\left( \sum {E}_R[f](t)^{2} \right)^{1/2}.
		\end{align}
		For each fixed $t$, $\mathcal{E}(t)$ provides a single number that can be used to quantity the total error in our numerical simulations.

\begin{figure}
			\centering
			\includegraphics[width=1.0\linewidth]{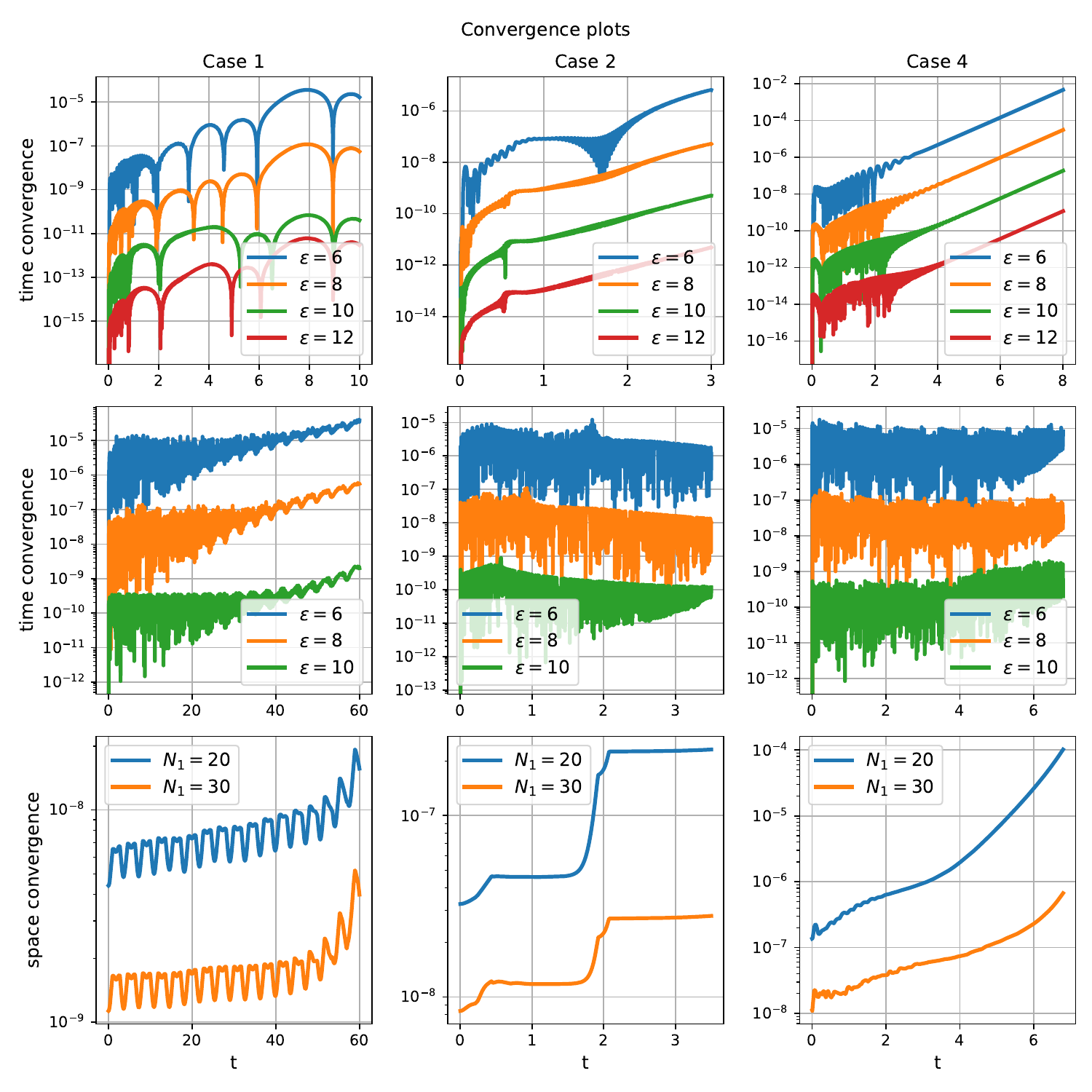}
			\caption{Plots demonstrating the convergence of our code. In regards to our model parameters, the first column shows simulations corresponding to the choices $(m_0,m_1,m_2,\omega)=(0.25,1,-1,1)$. Similarly, the second and third columns shows the results corresponding to the parameters $(m_0,m_1,m_2,\omega)=(0.5,2,1,1)$ and $(m_0,m_1,m_2,\omega)=(-0.234375,0.25,1,1)$, respectively. The first row shows convergence for spatially homogenous solutions with $N=1$. The second and third rows show the time and spatial convergence for spatially inhomogeneous solutions initial data with $\delta=10^{-3}$ (see \Eqref{Eq:MeanCurvePerturb}).}
			\label{fig:convergence}
		\end{figure}

		In \Figref{fig:convergence} we show convergence plots for six numerical simulations. In the first row of \Figref{fig:convergence} we show convergence for spatially homogenous solutions. In this case, we use our exact solutions as a reference solution. Here, Case 1, Case 2, and Case 3, correspond to the choices $(m_0,m_1,m_2,\omega)=(0.25,1,-1,1)$, $(m_0,m_1,m_2,\omega)=(0.5,2,1,1)$, and $(m_0,m_1,m_2,\omega)=(-0.234375,0.25,1,1)$, respectively. In the second and third rows of \Figref{fig:convergence} we demonstrate time and space convergence for spatially inhomogeneous perturbations of our exact solutions. Here, the initial data is constructed as described in Appendix~\ref{Appendix:InitialData} with $\delta=10^{-3}$. For all three of these cases, the reference solution is calculated with $\varepsilon=12$ and $N_2=40$.In all of these simulations, we find that the observed convergence is consistent with our numerical scheme. 
	
		\newpage
			
		\section{Summary of our findings}
		\label{Sec:Summary_of_our_findings}
		\begin{table}[t!]
			\centering
			\begin{tabular}{|c|c|c|c|c|}
				\hline
				Case & $t\rightarrow-\infty$ & $t\rightarrow+\infty$ & Stable 1: $t\rightarrow (-\infty,+\infty)$ & Stable 2: $t\rightarrow(-\infty,+\infty)$ \\
				\hline
				$1(\pm)$ & Oscillates & Oscillates & (Yes,Yes) & (No,No) \\
				\hline
				$2(+)$ & Big Bang & Big Crunch & (No,No) & (N/A,N/A) \\
				\hline
				$2(-)$ & Big Rip & Big Rip & (Yes,Yes) & (Yes,Yes) \\
				\hline
				$3(+)$ & Big Rip & Big Crunch & (Yes,No) & (Yes,N/A) \\
				\hline
				$3(-)$ & Big Bang & Big Rip & (No,Yes) & (N/A,Yes) \\
				\hline
				$4(-)$ & Big Bang & Static & (No,Yes) & (N/A,No) \\
				\hline
			\end{tabular}
			\caption{\label{Table:Paper_summary} A tabulated summary of the key findings of this paper, including a summary of the solutions stability properties, under mean curvature perturbation.}
		\end{table}
		In this work we have constructed four exact solutions of the Einstein equations coupled to a non-variational ``near-minimal'' scalar field, and investigated their stability properties. A summary of our findings is provided in Table~\ref{Table:Paper_summary}. In column $1$, of Table~\ref{Table:Paper_summary}, we provide the case number (corresponding the ordering of the solutions given in \Sectionref{Sec:Exact_solutions}). Here the $(\pm)$, next to the case number, represents the sign of $\partial_{t}\phi$ which must be chosen when integrating \Eqref{Eq:phi_quad}. Columns $2$ and $3$ summarise the behaviour of the solutions in the limits $t\rightarrow-\infty$ and $t\rightarrow+\infty$, respectively. Column $4$ states whether or not the solutions are stable to spatially homogenous mean curvature perturbations in the limits $t\rightarrow\pm\infty$. e.g., (Yes, No) means that, under perturbations of this type, the solution is stable as $t\rightarrow-\infty$ and unstable as $t\rightarrow\infty$. Column $5$ states whether or not the solutions are stable to spatially inhomogeneous mean curvature perturbations in the limits $t\rightarrow\pm\infty$. Note that, if the solutions were found to be unstable in the spatially homogenous setting, then their stability was not tested in the spatially inhomogeneous setting. In these cases we write ``N/A'' meaning ``not applicable''.

	\end{appendices}
\end{document}